\documentclass[useAMS,usenatbib,referee]{biom}

 \newcommand{\bitem}{\begin{itemize}}
 \newcommand{\eitem}{\end{itemize}}

 \newcommand{\R}{\mbox{\boldmath $R$}}

 \newcommand{\Z}{\mbox{\boldmath $Z$}}

\usepackage[figuresright]{rotating}
\usepackage{amsmath}

\usepackage{amssymb}

\let\arxivMathbb\mathbb
\renewcommand{\mathbb}[1]{\mathbf{#1}}

\providecommand{\mathbb}[1]{\mathbf{#1}}
\usepackage{url}
\usepackage{pifont}
\usepackage{booktabs}
\usepackage{algorithm}
\usepackage{algpseudocode}
\usepackage{xcolor}
\usepackage[
  colorlinks=true,
  hypertexnames=false,
  linkcolor=purple,
  citecolor=purple,
  urlcolor=purple,
  filecolor=purple,
  runcolor=purple
]{hyperref}

\volume{}
\pubyear{}
\artmonth{}
\doi{}
\pagerange{}
\journal{}
\makeatletter
\def\ps@headings{%
  \let\@mkboth\markboth
  \def\@oddhead{\small\itshape\@shorttitle\hfill\normalfont\thepage}%
  \def\@evenhead{\small\normalfont\thepage\hfill\itshape\@shorttitle}%
  \def\@oddfoot{}%
  \def\@evenfoot{}%
  \def\sectionmark##1{\markboth{##1}{}}%
  \def\subsectionmark##1{\markright{##1}}%
}
\let\submission@figurecaption\SFB@makefigurecaption
\long\def\SFB@makefigurecaption#1#2{%
  \begingroup
  \def\baselinestretch{1.8}\normalsize
  \submission@figurecaption{#1}{#2}%
  \endgroup
}
\long\def\SFB@maketablecaption#1#2{\vskip 6pt
  {\def\baselinestretch{1.8}\reset@font\normalsize\centering
    {\bf #1}\par\vskip1pt
    {\it #2}\par}%
  \vskip 0pt
  \@contfalse
}
\makeatother
\newcommand{\figurecaptionsheet}{\end{figure}\clearpage\begin{figure}\centering}

\makeatletter
\g@addto@macro\ps@titlepage{\let\@oddfoot\@empty\let\@evenfoot\@empty}
\makeatother

\title[Additive Gaussian Processes for Microbiome Data]{Bayesian Multinomial Logistic Normal Additive Regression for Microbiome Data Analysis}

\author{Tinghua Chen$^{1}$, 
Michelle Pistner Nixon$^{1}$, Lawrence A. David$^{2}$, and Justin D. Silverman$^{1,3,4,5*}$\email{JustinSilverman@psu.edu} \\
$^{1}$College of Information Sciences and Technology, Pennsylvania State University, University Park, PA, U.S.A. \\
$^{2}$ Center for Genomics and Computational Biology, Duke University, Durham, NC, U.S.A\\
$^{3}$ Department of Statistics, Pennsylvania State University, University Park, PA, U.S.A. \\
$^{4}$ Department of Medicine, Pennsylvania State University, Hershey, PA, U.S.A.\\
$^{5}$ Institute for Computational and Data Science, Pennsylvania State University, University Park, PA, U.S.A. \\
$^{*}$Correspondence: Justin D. Silverman, 514b Wartik Building, \\
360 Science Drive, University Park, PA 16802, U.S.A. }

\DeclareMathOperator*{\arxivSuppInf}{Inf}
\DeclareMathOperator*{\arxivSuppSup}{Inf}

\begin{document}

\label{m-firstpage}

\begin{abstract}
Longitudinal microbiome studies often seek to separate background temporal
variation from nonlinear perturbation-associated dynamics while accounting
for the count-compositional nature of sequencing data. We introduce
\textit{MultiAddGPs}, a semi-parametric Bayesian multinomial logistic-normal
framework that combines linear effects with additive Gaussian process
components. We first prove that MultiAddGPs remain within the Marginally Latent
Matrix-T Process class (MLTP) and derive the collapsed and uncollapsed representations
needed for scalable inference with the Collapse--Uncollapse (CU) sampler. We then
develop a back-sampling procedure for efficient posterior inference on the
individual additive components. We further develop approximate maximum marginal likelihood and
penalized maximum marginal likelihood procedures for 
hyperparameter estimation in this broader class of models. In simulations,
MultiAddGPs provides accurate posterior inference and calibrated uncertainty
while scaling more effectively than existing GP-based additive alternatives.
Applications to two longitudinal microbiome studies demonstrate how the model
separates background temporal variation from intervention-associated dynamics,
including travel-associated perturbations in a human gut time series and starvation effects
 in an artificial gut system. Our methods are implemented in the \textit{fido} package on CRAN.
\end{abstract}


\maketitle

\section{Introduction}
\label{m-sec: Introduction}

Microbiome studies increasingly aim to quantify how covariates such as diet, host physiology, environment, and time shape microbial community composition. These effects are often nonlinear and temporally structured, as demonstrated by~\citet{m-david2014host} in a year-long study tracking daily gut microbiome dynamics across two individuals. Their findings showed that while microbial communities are broadly stable, acute events—such as international travel or diarrheal illness—induce sharp, temporary shifts. This phenomenon highlights key analytical challenges: isolating perturbation-driven dynamics from background temporal variation, and identifying the specific taxonomic features that drive these compositional changes. Addressing these questions requires models that flexibly decompose longitudinal compositional dynamics while retaining interpretable summaries of covariate-associated effects ~\citep{m-cheng2019additivegp,m-sankaran2024mbtransfer,m-silverman2018dynamic}.

Additive regression models offer a natural framework for such decompositions. Common approaches include generalized additive models (GAMs), which represent
nonlinear covariate effects through smooth functions~\citep{m-wood2017generalized}, and longitudinal Gaussian process regression (Lgpr), which uses additive Gaussian process (GP) components to model longitudinal effects~\citep{m-timonen2021Lgpr}. Gaussian process (GP)-based additive models are especially appealing because they flexibly capture complex smooth relationships. However, existing GP-based additive approaches are computationally intensive or typically rely on Gaussian assumptions applied to transformed data, limiting their suitability for sequencing-based microbiome observations~\citep{m-cheng2019additivegp,m-timonen2021Lgpr}. This motivates scalable statistical methods that combine additive structure with models tailored to the measurement properties of microbiome data.

A central challenge is that microbiome measurements are obtained as a \(D \times N\) table of sequence counts \(\mathbf{Y}\), where \(Y_{dn}\) is the number of reads assigned to taxon \(d\) in sample \(n\). These counts are a random sample from an underlying microbial population, and the total count \(\lambda_n=\sum_{d=1}^{D}Y_{dn}\) (sequencing depth) is typically arbitrary and unrelated to absolute microbial load~\citep{m-vandeputte2017quantitative}. Consequently, sequencing data primarily identify relative abundances and are commonly treated as compositional~\citep{m-gloor2016compositional,m-mao2022dirichlet}. Unlike classical continuous compositional data, however, microbiome observations are zero-laden counts with substantial sampling noise, which complicates standard log-ratio pipelines and Gaussian error models~\citep{m-kaul2017analysis,m-Li2015microbiome}.
While multivariate zero-inflated models have been developed for sparse microbiome data~\citep{m-lee2018bayesian,m-shuler2021bayesian}, they can inappropriately infer that rarely observed taxa are highly abundant but unobserved because of high rates of zero inflation, a conclusion at odds with the sequencing measurement process~\citep{m-silverman2020naught}.

Bayesian multinomial logistic-normal (MLN) models address these challenges by combining a multinomial sampling model with a logistic-normal latent representation that permits rich covariance among taxa~\citep{m-aitchison1980,m-silverman2022bayesian}. Under this formulation, observed zeros can arise naturally through multinomial sampling, without requiring an additional zero-inflation component. A major computational advance was introduced by~\citet{m-silverman2022bayesian}, who developed the Marginally Latent Matrix-T Process (MLTP) framework and the Collapse-Uncollapse (CU) sampler for scalable inference in a broad class of MLN regression models, including either linear or Gaussian process regression. This framework provides the computational foundation for the present work.

However, three important gaps remain. First, the previous theory does not establish that a model formed from additive linear and Gaussian process components remains within the MLTP class, nor does it provide the collapsed and uncollapsed representations needed to apply the CU sampler to that hierarchy. Second, the existing framework does not recover the component processes as separately interpretable components. For longitudinal microbiome studies, such a decomposition is essential when background temporal variation, linear covariate effects, and intervention-associated nonlinear responses act simultaneously. Recovering these components requires posterior inference beyond the latent aggregate function. Third, the original MLTP framework treats model hyperparameters as fixed and therefore does not provide a general mechanism for estimating quantities such as kernel hyperparameters from the data.

In this article, we build on the MLTP framework of \citet{m-silverman2022bayesian} and make three methodological contributions that substantially expand its inferential scope. First, we establish that MultiAddGPs remain within the MLTP class despite their additive structure, and we derive the parameters of the collapsed LTP representation and the corresponding uncollapsed conditional representation. This result is what makes scalable CU inference possible for the new model class. Second, we develop a back-sampling procedure that efficiently reconstructs component-specific posterior draws from the aggregate posterior. This procedure recovers multiple overlapping processes---including linear effects, background temporal variation, and nonlinear intervention responses---as separately interpretable components. Third, we develop computationally efficient approximate maximum marginal likelihood and penalized maximum marginal likelihood procedures for hyperparameter estimation in MLTP models. This contribution is not specific to MultiAddGPs: our approach applies to any model within the larger MLTP class. Together, these developments transform the MLTP framework from a scalable engine for additive regression with fixed hyperparameters into a practical framework for learning and interpreting multiple component processes in count-compositional data. This substantially broadens the scientific utility of MLTP models. In longitudinal microbiome studies, for example, additive decomposition makes it possible to isolate travel- or starvation-associated perturbations from background temporal variation and to identify the taxa driving those responses. Table~\ref{m-tb:model_comparison} summarizes how these capabilities distinguish MultiAddGPs from the existing MLTP framework and other additive regression approaches.

\begin{table}[H]
    \centering
    \resizebox{\textwidth}{!}{%
    \begin{tabular}{lcccc}
        \toprule
        & \textit{MultiAddGPs} & \textit{MLTPs} & \textit{Lgpr}  & \textit{GAMs}  \\
        \midrule
      Flexible Kernel                 & \ding{51} & \ding{51} & \ding{53} & NA \\
        Scalable                          & \ding{51} & \ding{51} & \ding{53} & \ding{51} \\
        Hyperparameter Estimation         & \ding{51} & \ding{53} & \ding{51} & \ding{51} \\
        Count Compositional               & \ding{51} & \ding{51} & \ding{53} & \ding{53} \\
        Effect Decomposition              & \ding{51} & \ding{53} & \ding{51} & \ding{51} \\
        \bottomrule
    \end{tabular}%
    }
    \caption{\textbf{A comparison of MultiAddGPs and existing alternatives.} MultiAddGPs (this article) facilitate additive modeling of both linear and nonlinear processes. In contrast, the MLTP models presented by \protect\citet{m-silverman2022bayesian} contain either a single linear or a single nonlinear component and do not provide a mechanism for decomposing mixed components. Lgpr~\protect\citep{m-timonen2021Lgpr} provides additive Gaussian process modeling. Both MultiAddGPs and MLTPs are implemented in the \textit{fido} package available on CRAN and support predefined and user-defined kernels, whereas Lgpr uses predefined kernels. MultiAddGPs and MLTPs use the computationally efficient CU sampler; Lgpr was computationally intensive and could be evaluated only for a subset of the simulation settings in this article. The original MLTP formulation treats model hyperparameters as fixed, whereas the other three methods provide hyperparameter or smoothing-parameter estimation. MultiAddGPs and MLTPs address count-compositional observations using Bayesian MLN models, whereas Lgpr uses univariate likelihoods. GAMs provide additive effect decomposition through smooth terms but, as implemented in our comparison, do not jointly model count-compositional observations. NA denotes that kernel flexibility is not applicable to the spline-based GAMs considered here.}
    \label{m-tb:model_comparison}
\end{table}

We also evaluate MultiAddGPs through simulation studies and applications to two longitudinal microbiome datasets. These analyses show that MultiAddGPs yields calibrated uncertainty and scalable computation while enabling interpretable decompositions of background temporal variation and intervention-associated dynamics, including travel-associated perturbations in a human gut time series and starvation effects in an artificial gut system. 

The remainder of this article is organized as follows. Section~\ref{m-sec: methods} introduces the MultiAddGP framework and develops posterior inference, component-wise back-sampling, model identification, and marginal-likelihood-based hyperparameter estimation. Section~\ref{m-sec:simulation} evaluates the statistical and computational performance of MultiAddGPs through simulation studies. Section~\ref{m-sec:longitudinal} presents the two longitudinal microbiome applications, and Section~\ref{m-sec:discussion} concludes with a discussion of limitations and future directions.

\vspace{-0.5cm}
\subsection*{Notation}

\label{m-sec:notation}

We use unbolded uppercase letters (e.g., \(N\)) for dimensions, bold uppercase symbols (e.g., \(\boldsymbol{X}\)) for matrices, bold lowercase symbols (e.g., \(\mathbf{x}\)) for vectors, and unbolded lowercase symbols (e.g., \(x\)) for scalars. Scalar entries of matrices are unbolded and retain the corresponding matrix symbol (e.g., \(Y_{dn}\)). Matrix rows and columns are indexed as \(\mathbf{X}_{d\cdot}\) and \(\mathbf{X}_{\cdot n}\), respectively. For a vector-valued
function \(\mathbf{f}^{(k)}\), we denote its evaluations across \(N\) observations by the matrix $ \boldsymbol{F}^{(k)} =\left[ \mathbf{f}^{(k)}(\boldsymbol{Z}^{(k)}_{\cdot 1}),\ldots, \mathbf{f}^{(k)}(\boldsymbol{Z}^{(k)}_{\cdot N}) \right],$ so that \(\boldsymbol{F}^{(k)}_{\cdot n}
= \mathbf{f}^{(k)}(\boldsymbol{Z}^{(k)}_{\cdot n})\).
\vspace{-0.5cm}

\section{Methods}
\label{m-sec: methods}

\subsection{Multinomial Logistic Normal Additive Gaussian Process Models (MultiAddGPs)} 
\label{m-sec: multiaddgps}

For sample \(n=1,\ldots,N\), let \(\mathbf{Y}_{\cdot n}\) denote a
\(D\)-vector of observed counts, \(\mathbf{X}_{\cdot n}\) a
\(Q_0\)-vector of covariates modeled linearly, and
\(\mathbf{Z}^{(k)}_{\cdot n}\) a \(Q_k\)-vector of covariates modeled
nonlinearly for \(k=1,\ldots,K\). Let
\(\boldsymbol{\Pi}_{\cdot n}\in\mathcal{S}^{D}\) denote the corresponding
composition, and let
\(\mathbf{H}_{\cdot n},\mathbf{F}_{\cdot n}\in\mathcal{R}^{D-1}\)
denote the latent log-ratio state and its mean, respectively.
Stacking these quantities across samples gives
\(\mathbf{X}\in\mathcal{R}^{Q_0\times N}\),
\(\mathbf{Z}^{(k)}\in\mathcal{R}^{Q_k\times N}\), and
\(\mathbf{H},\mathbf{F}\in\mathcal{R}^{(D-1)\times N}\). Finally, let \(\lambda_n=\sum_{d=1}^D Y_{dn}\) denote the total count, or sequencing depth, for sample \(n\). MultiAddGPs have the form:
\begin{align}
  \mathbf{Y}_{\cdot n} &\sim \text{Multinomial}(\lambda_n,\boldsymbol{\Pi}_{\cdot n}), \label{m-eqs:multinomial}\\
  \boldsymbol{\Pi}_{\cdot n} &= \phi^{-1}(\mathbf{H}_{\cdot n}), \label{m-eqs:logratiotransform}\\
  \mathbf{H}_{\cdot n} &\sim  \mathcal{N}(\mathbf{F}_{\cdot n}, \boldsymbol{\Sigma}),
  \label{m-eqs:State_F}\\
  \mathbf{F} &= \mathbf{B}\mathbf{X} + \sum_{k=1}^{K}\mathbf{F}^{(k)}. \label{m-eqs:F}
\end{align}

Here, $\mathbf{B}\in \mathcal{R}^{(D-1)\times Q_0}$. Each column of \(\mathbf{F}^{(k)}\) is the value of the vector-valued function \(\mathbf{f}^{(k)}\) at the corresponding sample's covariates. We assign priors $\mathbf{B}\sim\mathcal{MN}(\boldsymbol{\Theta}^{(0)},\boldsymbol{\Sigma},\boldsymbol{\Gamma}^{(0)})$, $\mathbf{f}^{(k)}\sim\mathcal{GP}(\boldsymbol{\Theta}^{(k)},\boldsymbol{\Sigma},\boldsymbol{\Gamma}^{(k)})$, and $\boldsymbol{\Sigma}\sim\mathcal{IW}(\boldsymbol{\Xi},\zeta)$, where $\mathcal{MN}$ denotes the matrix-normal distribution, $\mathcal{GP}$ denotes a multivariate Gaussian process, and $\mathcal{IW}$ denotes the inverse-Wishart distribution, with $\boldsymbol{\Xi}$ denoting its scale matrix and $\zeta$ its degrees-of-freedom
parameter, with $\zeta>D-2$. The distributional conventions and densities are given in~\ref{s-sec:distribution-conventions}. Here, \(\boldsymbol{\Sigma}\) is a \((D-1) \times (D-1)\) matrix representing covariance between taxonomic log-ratios, and the transform \(\phi\) is an invertible log-ratio transform from the \(D\)-dimensional simplex to \(D-1\) dimensional real-space: \(\mathbf{H}_{\cdot n}  = \phi(\boldsymbol{\Pi}_{\cdot n}) \in \mathcal{R}^{D-1}\). For computational efficiency, we used the following Additive Log-Ratio (ALR) transform, which takes the \(D\)-th taxa as a reference:
\begin{equation*}
\mathbf{H}_{\cdot n} = \phi(\boldsymbol{\Pi}_{\cdot n}) =  \left\{ \log\left(\frac{\pi_{1n}}{\pi_{Dn}}\right), \ldots, \log\left(\frac{\pi_{(D-1)n}}{\pi_{Dn}}\right) \right\}^{T}.
\end{equation*}
There is no loss of generality as posterior samples taken with respect to the \(\text{ALR}_{D}\) coordinate system can be transformed into any other log-ratio coordinate system~\citep*[Appendix A.3]{m-pawlowsky2015modeling}. 

For the matrix-normal prior on the linear term, \(\boldsymbol{\Theta}^{(0)}\) denotes the mean matrix and \(\boldsymbol{\Gamma}^{(0)}\) a \(Q_0 \times Q_0\) covariance matrix governing dependence among regression coefficients. For \(k=1,\ldots,K\), the corresponding quantities \(\boldsymbol{\Theta}^{(k)}\) and \(\boldsymbol{\Gamma}^{(k)}\) define mean and kernel functions for the matrix-normal process priors on the nonlinear components. We collect all kernel and mean-function hyperparameters into \(\Omega\).

\subsection{Posterior Estimation in MultiAddGPs}
\label{m-sec:multiaddgps-inference}

Posterior inference is performed using the Collapse--Uncollapse (CU) sampler framework of \citet{m-silverman2022bayesian}.The sampler proceeds in two stages: the \emph{collapse} step samples from the marginal posterior \(p(\mathbf{H}\mid\mathbf{Y})\), which in MLTP models corresponds to a latent matrix-T process (LTP) and is approximated using a Laplace approximation; the \emph{uncollapse} step then samples from the conditional posterior \(p(\boldsymbol{\Phi}\mid\mathbf{H},\mathbf{Y})\), where $\boldsymbol{\Phi}$ denotes the collection of model parameters, which admits a closed form due to conjugacy. This enables scalable inference for MultiAddGPs. A detailed review of the CU sampler and MLTP formulation is provided in~\ref{s-sec: Marginally-LTP}.

In the following Section, we first show that MultiAddGPs belong to the MLTP class and derive their collapsed LTP representation, which determines the Laplace approximation used in the collapse step. We then develop a back-sampling procedure that yields closed-form updates for the uncollapse step in multivariate additive Gaussian process models. Finally, we describe marginal likelihood–based hyperparameter estimation within the MLTP framework and discuss identification of posterior samples. Algorithm~\ref{m-algorithm:CU} summarizes the inference algorithm (We omit conditioning on covariates $\mathbf{X}$ and $\mathbf{Z}$ in the notation for brevity.). 

\begin{algorithm}[t]
    \caption{The Collapse-Uncollapse (CU) Sampler for \textbf{MultiAddGPs} Models}
    \label{m-algorithm:CU}
    \resizebox{\textwidth}{!}{%
    \parbox{\textwidth}{%
        \raggedright
        \textbf{Input:} Data $\{\mathbf{Y}, \mathbf{X}, \mathbf{Z}^{(1)}, \dots, \mathbf{Z}^{(K)}\}$, and prior parameters $\Delta = \{\boldsymbol{\Theta}^{(0)},\dots,\boldsymbol{\Theta}^{(K)}, \boldsymbol{\Gamma}^{(0)},\dots,\boldsymbol{\Gamma}^{(K)}, \boldsymbol{\Xi}, \zeta\}$.\\
        \textbf{Output:} $S$ samples of $\{\mathbf{H}, \boldsymbol{\Sigma}, \mathbf{F}, \mathbf{B}, \mathbf{F}^{(1)}, \dots,\mathbf{F}^{(K)}\}$. 
        \begin{algorithmic}[1]
          \For {each sample \(s \in \{1, \dots, S\}\)}
            \State Sample $\mathbf{H}^{[s]} \sim p(\mathbf{H}|\mathbf{Y},\Delta)$ where $p(\mathbf{H},\mathbf{Y},\Delta)$ is an LTP;
            \Comment{Collapsed Step}
            \State Sample \(\{\boldsymbol{\Sigma}^{[s]}, \mathbf{F}^{[s]}\} \sim p(\mathbf{F}, \boldsymbol{\Sigma} \mid  \mathbf{H}^{[s]},\Delta)\);
            \Comment{Uncollapsed Step (1)}
            \State Sample $\mathbf{B}^{[s]}, \mathbf{F}^{(1)[s]}, \dots, \mathbf{F}^{(K)[s]}$ $\sim \text{Back-Sampling}(\mathbf{Y}, \mathbf{F}^{[s]}, \boldsymbol{\Sigma}^{[s]},\Delta)$;
            \Statex \Comment{Uncollapsed Step (2)}
          \EndFor
        \end{algorithmic}
    }
    }
\end{algorithm}

\subsubsection{MultiAddGPs as MLTPs and the Collapse Step}
\label{m-subsec: multiaddgps-collpasedstep}
Here, \(\mathbf{Y}\), \(\boldsymbol{\Pi}\), and \(\mathbf{H}\) denote the
corresponding quantities stacked across all \(N\) samples. A Latent Matrix-T Process (LTP) is a hierarchical model of the form
\begin{align*}
\mathbf{Y} &\sim g(\boldsymbol{\Pi}), \\
  \boldsymbol{\Pi}&=\phi^{-1}(\mathbf{H}), \\
  \mathbf{H} &\sim TP(\nu, \mathbf{M}, \mathbf{V}, \mathbf{A}), 
\end{align*}
where \(\phi\) is a known transform and \(TP(\nu,\mathbf{M},\mathbf{V},\mathbf{A})\) denotes a matrix-T process with location function \(\mathbf{M}\), row scale \(\mathbf{V}\), column scale \(\mathbf{A}\), and degrees of freedom \(\nu\) \citep{m-silverman2022bayesian}. A model is Marginally LTP (MLTP) if its joint distribution admits this representation after marginalizing over auxiliary parameters. In \ref{s-sec:proof}, we prove the following theorem, which establishes that MultiAddGPs are MLTPs.

\begin{theorem}
  MultiAddGPs are MLTPs with parameters  $\boldsymbol{\Phi} = \{\mathbf{F},\mathbf{B}, \mathbf{F}^{(1)}, \dots, \mathbf{F}^{(K)}, \boldsymbol{\Sigma}\}$. The marginal form \(p( \mathbf{Y},\mathbf{H})\) is an LTP:
  \begin{align*}
  \mathbf{Y}_{\cdot n} &\sim \text{Multinomial}(\lambda_n,\boldsymbol{\Pi}_{\cdot n}),  \\ 
    \boldsymbol{\Pi}_{\cdot n} &= \phi^{-1}(\mathbf{H}_{\cdot n}), \\
    \mathbf{H} &\sim TP(\nu, \mathbf{M}, \mathbf{V}, \mathbf{A}),
  \end{align*}
 where $\phi$ is an invertible log-ratio transformation, $  \mathbf{M}=\boldsymbol{\Theta}^{(0)}\mathbf{X} + \sum_{k=1}^K\boldsymbol{\Theta}^{(k)}(\mathbf{Z}^{(k)})$, $\mathbf{V} = \boldsymbol{\Xi}$, and $\mathbf{A} = \mathbf{X}^{T} \boldsymbol{\Gamma}^{(0)} \mathbf{X} + \sum_{k=1}^{K} \boldsymbol{\Gamma}^{(k)}(\mathbf{Z}^{(k)}) + \mathbf{I}_N, \nu =\zeta-D+2$.
 \end{theorem}

Let $\operatorname{vec}(\mathbf{H})$ denote the vector obtained by stacking
the columns of $\mathbf{H}$. In the collapsed step of the CU sampler, we approximate \(p(\mathbf{H} \mid \mathbf{Y})\) using a Laplace approximation: $q(\mathbf{H}| \mathbf{Y}) = \mathcal{N}\left(
\operatorname{vec}(\widehat{\mathbf{H}}),
\left[\nabla^{2}\{\operatorname{vec}(\widehat{\mathbf{H}})\}\right]^{-1}
\right)$,
where \(\widehat{\mathbf{H}}\) is the maximum a posteriori (MAP) estimate of \(p(\mathbf{H} \mid \mathbf{Y})\), and \(\left[\nabla^{2}\{\operatorname{vec}(\widehat{\mathbf{H}})\}\right]^{-1}\) is the inverse Hessian of negative log-posterior evaluated at \(\widehat{\mathbf{H}}\). We obtain efficient MAP estimates $\widehat{\mathbf{H}}$ by applying the L-BFGS optimizer to the 
closed-form gradients $\nabla\{\mathrm{vec}(\widehat{\mathbf{H}})\}$ derived in 
\ref{s-sec:LA}. Building on these results, we then compute the Hessian 
$\nabla^{2}\{\mathrm{vec}(\widehat{\mathbf{H}})\}$ in closed form (also see \ref{s-sec:LA}). This approximation is computationally efficient and accurate. The error in this approximation is bounded in probability by  \(O_{p}\left\{(D-1)\sum_{n=1}^{N} \lambda_{n}^{-1}\right\}\) (See Proposition 11 of \cite{m-silverman2022bayesian}), implying improved
accuracy at larger sequencing depths. \citet{m-silverman2022bayesian} also
showed that the resulting CU sampler was often 4--5 orders of magnitude
faster than MCMC in their empirical evaluations. Our simulations extend this result: despite jointly modeling the multivariate composition, MultiAddGPs was orders of magnitude faster than fitting separate single-response Lgpr models using Hamiltonian Monte Carlo (Section~\ref{m-sec:simulation}; Figure~\ref{m-fig:runtime}).

\subsubsection{Back-sampling and the Uncollapse Step for MultiAddGPs}
\label{m-subsec:multiaddgps-backsampler}

Given a posterior draw \(\mathbf{H}\sim p(\mathbf{H}\mid\mathbf{Y})\), the uncollapse step samples 
\(\boldsymbol{\Phi}=\{\mathbf{F},\mathbf{B},\mathbf{F}^{(1)},\dots,\mathbf{F}^{(K)},\boldsymbol{\Sigma}\}\) 
from \(p(\boldsymbol{\Phi}\mid\mathbf{H},\mathbf{Y})\). Since 
\(\boldsymbol{\Phi}\perp\mathbf{Y}\mid\mathbf{H}\), this reduces to sampling 
\(p(\boldsymbol{\Phi}\mid\mathbf{H})\). Moreover, the conditional independence relation \(\{\mathbf{B}, \mathbf{F}^{(1)}, \dots, \mathbf{F}^{(K)}\} \perp \mathbf{H} \mid  \{\mathbf{F}, \boldsymbol{\Sigma}\}\) implies that the uncollapse step can be decomposed into two simpler steps: first sample \(p(\boldsymbol{\Sigma}, \mathbf{F} \mid  \mathbf{H})\) and then sample \(p(\mathbf{B}, \mathbf{F}^{(1)},\dots,  \mathbf{F}^{(K)} \mid  \mathbf{F}, \boldsymbol{\Sigma})\). Thus
\[
p(\boldsymbol{\Phi}\mid\mathbf{H})
=
p(\boldsymbol{\Sigma},\mathbf{F}\mid\mathbf{H})\,
p(\mathbf{B},\mathbf{F}^{(1)},\dots,\mathbf{F}^{(K)}\mid\mathbf{F},\boldsymbol{\Sigma}).
\]

\paragraph{Notation:} We distinguish between the observed sample indices \(n\in \{1,\dots,N\}\), where \(\mathbf{Y}\) and \(\mathbf{H}\) are defined, and the set of input locations at which the additive components are evaluated \(n^{*} \in \{1, \dots, N^{*}\}\). In what follows, \(\mathbf{F},\mathbf{F}^{(1)},\ldots,\mathbf{F}^{(K)}\) denote matrices of evaluations at the chosen input locations, while \(\mathbf{B}\) denotes the linear coefficient matrix.

\paragraph{Sampling \(\boldsymbol{\Sigma},\mathbf{F}\mid\mathbf{H}\)}
Conditioning on \(\mathbf{H}\) and marginalizing over \(\mathbf{B}\) and \(\mathbf{F}^{(k)}, k \in \{1,\dots, K\}\) in the MultiAddGPs model results in a Bayesian matrix-normal process model with likelihood \(\mathbf{H}_{\cdot n} \sim \mathcal{N}(\mathbf{F}_{\cdot n},\boldsymbol{\Sigma})\) and priors:
\begin{align*}
  \mathbf{F} &\sim  \mathcal{MN}\left(\boldsymbol{\Theta}^{(0)}\mathbf{X} + \sum_{k=1}^K\boldsymbol{\Theta}^{(k)}(\mathbf{Z}^{(k)}),\; \boldsymbol{\Sigma},\; \mathbf{X}^{T}\boldsymbol{\Gamma}^{(0)}\mathbf{X} + \sum_{k=1}^{K}\boldsymbol{\Gamma}^{(k)}(\mathbf{Z}^{(k)})\right), \\ 
 \boldsymbol{\Sigma} &\sim \mathcal{IW}(\boldsymbol{\Xi}, \zeta).
\end{align*}
\noindent This model is conjugate and posterior samples of \(\mathbf{F}\) and \(\boldsymbol{\Sigma}\) can be efficiently obtained as described in \ref{s-subsec:sample-F-Sigma}.

\paragraph{Back-sampling \(\mathbf{B},\mathbf{F}^{(1)},\dots,\mathbf{F}^{(K)}\)}
Given \(\mathbf{F}\) and \(\boldsymbol{\Sigma}\), the parameters 
\(\mathbf{B}, \mathbf{F}^{(1)}, \dots, \mathbf{F}^{(K)}\) enter only through the additive decomposition
\(
\mathbf{F} = \mathbf{B}\mathbf{X} + \sum_{k=1}^{K} \mathbf{F}^{(k)}.
\)
We sample these components by decomposing \(\mathbf{F}\) into its linear and nonlinear contributions. 
To do so, we introduce a back-sampling procedure, conceptually related to backfitting in generalized additive models~\citep{m-wood2017generalized}, that sequentially draws each component from its full conditional distribution while conditioning on the remaining components.

Because the priors on \(\mathbf{B}\) and \(\mathbf{F}^{(k)}\) are matrix-normal, the resulting conditional distributions remain matrix-normal and admit closed-form updates. The algorithm first samples \(p(\mathbf{B}\mid \mathbf{F},\boldsymbol{\Sigma})\), then sequentially samples 
\[
p(\mathbf{F}^{(1)}\mid \mathbf{B},\mathbf{F},\boldsymbol{\Sigma}),\;
\dots,\;
p(\mathbf{F}^{(K-1)}\mid \mathbf{F}^{(1)},\dots,\mathbf{F}^{(K-2)},\mathbf{B},\mathbf{F},\boldsymbol{\Sigma}),
\]
after which, the final component is determined by the additive constraint
\[
\mathbf{F}^{(K)}
=
\mathbf{F}
-
\mathbf{B}\mathbf{X}
-
\sum_{k=1}^{K-1}\mathbf{F}^{(k)}.
\]
Closed-form expressions for the conditional distributions and pseudocode are given in Supplement~\ref{s-sec:backsampling}. The precision-form updates require positive-definite component covariance matrices; the numerical regularization used for gated kernels is described in~\ref{s-subsec:artificial_data}.

\subsection{Model Identification}
\label{m-subsec:method-identification}

Additive decomposition models such as MultiAddGPs are not fully identifiable without additional constraints, since different allocations of variation among components can yield the same fitted function~\citep{m-hastie2017generalized}. A common identification strategy is therefore to impose a sum-to-zero constraint on each nonlinear component, for example,
$\int \mathbf{f}^{(k)} \, d\mathbf{Z}^{(k)} = \mathbf{0}$.
When greater separation among flexible components is required, identification can also be strengthened by modifying the kernel functions to enforce orthogonality among components~\citep*{m-lu2022additive}. Both approaches are compatible with MultiAddGPs.

In this work, we approximate the sum-to-zero constraint by centering posterior samples of each component:
$\widetilde{\mathbf{f}}^{(k)}(\mathbf{z}^{(k)})=\mathbf{f}^{(k)}(\mathbf{z}^{(k)})-(1/N)\sum_{n=1}^{N}\mathbf{f}^{(k)}(\mathbf{Z}_{\cdot n}^{(k)}).$
When the evaluation points sufficiently cover the covariate space, this approximation closely enforces the desired constraint while remaining computationally simple.

\subsection{Inferring Model Hyperparameters}
\label{m-subsec:method-hyperparameters}

Let \(\Omega\) denote the collection of hyperparameters governing the MLTP model, including kernel, mean-function, and transformation parameters. We estimate \(\Omega\) via maximum marginal likelihood (MML),
\begin{align*}
\widehat{\Omega}
=
\arg\max_{\Omega}
\int p(\mathbf{H},\boldsymbol{\Phi},\mathbf{Y}\mid\Omega)
\, d\boldsymbol{\Phi}\, d\mathbf{H}.
\end{align*}
Because the model is MLTP, integrating out \(\boldsymbol{\Phi}\) yields the collapsed form \(p(\mathbf{H},\mathbf{Y}\mid\Omega)\), so the objective reduces to the marginal likelihood: $\log p(\mathbf{Y}\mid\Omega) = \log \int p(\mathbf{H},\mathbf{Y}\mid\Omega)\, d\mathbf{H}$.
We approximate this integral using the same Laplace approximation employed in the collapse step. Let \(\widehat{\mathbf{H}}_{\Omega}\) denote the MAP estimate under hyperparameters \(\Omega\). A first-order Laplace approximation gives
\begin{equation*}
  \log p(\mathbf{Y}\mid\Omega) \approx \frac{(D-1)N}{2}\log(2\pi) +  \log p(\widehat{\mathbf{H}}_{\Omega}, \mathbf{Y} \mid  \Omega)-\frac{1}{2} \log \left|\nabla^2\{\operatorname{vec}(\widehat{\mathbf{H}}_{\Omega})\}\right|.
\end{equation*}
Since the MAP estimate and Hessian are already computed during posterior inference, evaluating this approximation incurs little additional cost. 

Even after imposing the component-wise centering constraint described in Section~\ref{m-subsec:method-identification}, additive GP components may be difficult to distinguish when their kernels can represent similar patterns of variation. Prior knowledge can be incorporated into hyperparameter estimation by constraining the MML search space or adding a penalty to its objective. For example, suppose two temporal components use squared-exponential kernels
\(K_k(t,t')=\sigma_k^2\exp\{-(t-t')^2/(2\rho_k^2)\}\), where \(\rho_k\)
controls temporal scale and \(\sigma_k\) controls amplitude. 
If background variation is expected to be smoother than a transient perturbation,
the ordering $\rho_{\mathrm{background}}>\rho_{\mathrm{perturbation}}$ can be
imposed during optimization. Similarly,
$\sigma_{\mathrm{background}}<\sigma_{\mathrm{perturbation}}$ can be imposed
if the perturbation is expected to have a larger amplitude. Sensitivity can be assessed by
refitting under alternative constraints or penalty strengths.

Numerically, the resulting MML or penalized-MML objective is generally non-convex,
particularly with respect to kernel length-scale parameters. We therefore use
derivative-free Bayesian optimization, which is practical for the low- to
moderate-dimensional hyperparameter problems considered here. Here, 
optimization was performed using the \textit{ParBayesianOptimization} R package
with an upper-confidence-bound acquisition function, 10 initial points, and
20 optimization iterations; all other settings were left at their default values.

\subsection{Software Availability}
\label{m-subsec: methods-software}

MultiAddGPs is available through the \texttt{basset} function in the R package \texttt{fido} (version 1.1.5)~\citep{m-silverman2022bayesian}.  The \texttt{fido} package implements the extended CU sampler described above using optimized C++ code.

\section{Simulations}
\label{m-sec:simulation}

We evaluated MultiAddGPs under varying numbers of taxa \(D \in \{3, 10,50,100,300,500\}\) and sample sizes
\(N \in \{20, 50,70,100,300,500,1000\}\).
We compared against three representative alternatives.
First, we evaluated the additive Gaussian process regression framework implemented in the \textit{Lgpr} software \citep{m-timonen2021Lgpr}, fit under a Negative Binomial likelihood commonly used for microbiome count data \citep{m-zhang2017negative}.
Because microbiome counts are often normalized and analyzed on a log-ratio scale \citetext{\citealp*{m-kodikara2022statistical}; \citealp{m-worsley2024longitudinal}}, we also evaluated two Gaussian baselines applied to centered log-ratio (CLR) transformed outcomes $\mathrm{CLR}(\mathbf{Y}+0.65)$: Gaussian Lgpr and a GAM implemented using the \textit{mgcv} package. 

To evaluate recovery of overlapping temporal effects in the presence of a 
nonstationary perturbation, we simulated count-compositional data from
\[
    \mathbf{F}_{\cdot n}
    =
    \mathbf{b}_0
    +
    \mathbf{b}_1 x_n^{(\mathrm{Batch})}
    +
    \mathbf{f}^{(\mathrm{General})}(t_n)
    +
    \mathbf{f}^{(\mathrm{Intervention})}(w_n).
\]
Given this mean, we drew
$\mathbf{H}_{\cdot n}\sim\mathcal{N}(\mathbf{F}_{\cdot n},\boldsymbol{\Sigma})$
and set
$\boldsymbol{\Pi}_{\cdot n}=\operatorname{ALR}^{-1}(\mathbf{H}_{\cdot n})$.
Finally, we drew $\mathbf{Y}_{\cdot n}$ from a multinomial distribution with
size $\lambda_n$ and probability vector $\boldsymbol{\Pi}_{\cdot n}$, where
$\mathbf{F}_{\cdot n},\mathbf{H}_{\cdot n}\in\mathbb{R}^{D-1}$.
The terms $\mathbf{b}_0$ and $\mathbf{b}_1$ represent baseline and batch
effects, while $\mathbf{f}^{(\mathrm{General})}$ and
$\mathbf{f}^{(\mathrm{Intervention})}$ represent background temporal
variation and a localized intervention effect, respectively. We set
$\mathbf{b}_0=0.4$ and $\mathbf{b}_1=1$, and used a taxon covariance
matrix $\boldsymbol{\Sigma}$ with diagonal entries 1 and off-diagonal
entries 0.5.

The two nonlinear components were generated as mean-zero matrix-normal
processes. We used $\Gamma^{(\mathrm{General})}(t_n,t_{n'})=\exp\left\{-(t_n-t_{n'})^2/(2\times0.5^2)\right\}$ for the general component and $\Gamma^{(\mathrm{Intervention})}(w_n,w_{n'})
=4\exp\left\{-(w_n-w_{n'})^2/(2\times3^2)\right\}$ for the intervention component; both kernels are squared exponential.
Here, $w_n=w(t_n)$ is a skewed-Gaussian input transformation that induces a
localized, asymmetric perturbation centered near $t=20$. Thus, the
simulation combines a broad smooth temporal process with a shorter-lived
nonstationary perturbation acting on the same multivariate composition.
The explicit warping function and remaining data-generation details are
given in~\ref{s-subsec:sim-lgpr-gam}.

All methods were provided covariates corresponding to the batch, background temporal,
and intervention effects. MultiAddGPs modeled the background and intervention effects
using a squared-exponential kernel and a nonstationary kernel, respectively. Importantly,
the associated kernel hyperparameters were not fixed at the generating values, but were
estimated separately for each simulated dataset by Bayesian optimization of the penalized
log marginal likelihood described in Section~\ref{m-subsec:method-hyperparameters}. Lgpr
modeled the background temporal effect using the same squared-exponential temporal GP
and the intervention effect using its nonstationary GP component, together with a
categorical batch effect, under both negative binomial and Gaussian observation models.
GAM included a categorical batch effect, a smooth function of time for the background
temporal effect, and an adaptive smooth for the intervention effect.

For each \((D,N)\) configuration, 100 replicate datasets were generated.
Each method was allowed up to 48 hours of wall-clock runtime per replicate on a single compute node.
In practice, this limit primarily affected the Lgpr models, which were therefore evaluated only on a subset of the ($D,N$) settings. The exact configurations and implementation details, including prior specifications and model structure, are provided in~\ref{s-subsec:sim-lgpr-gam}.

\begin{figure}
    \centering
    \includegraphics[width=0.94\linewidth]{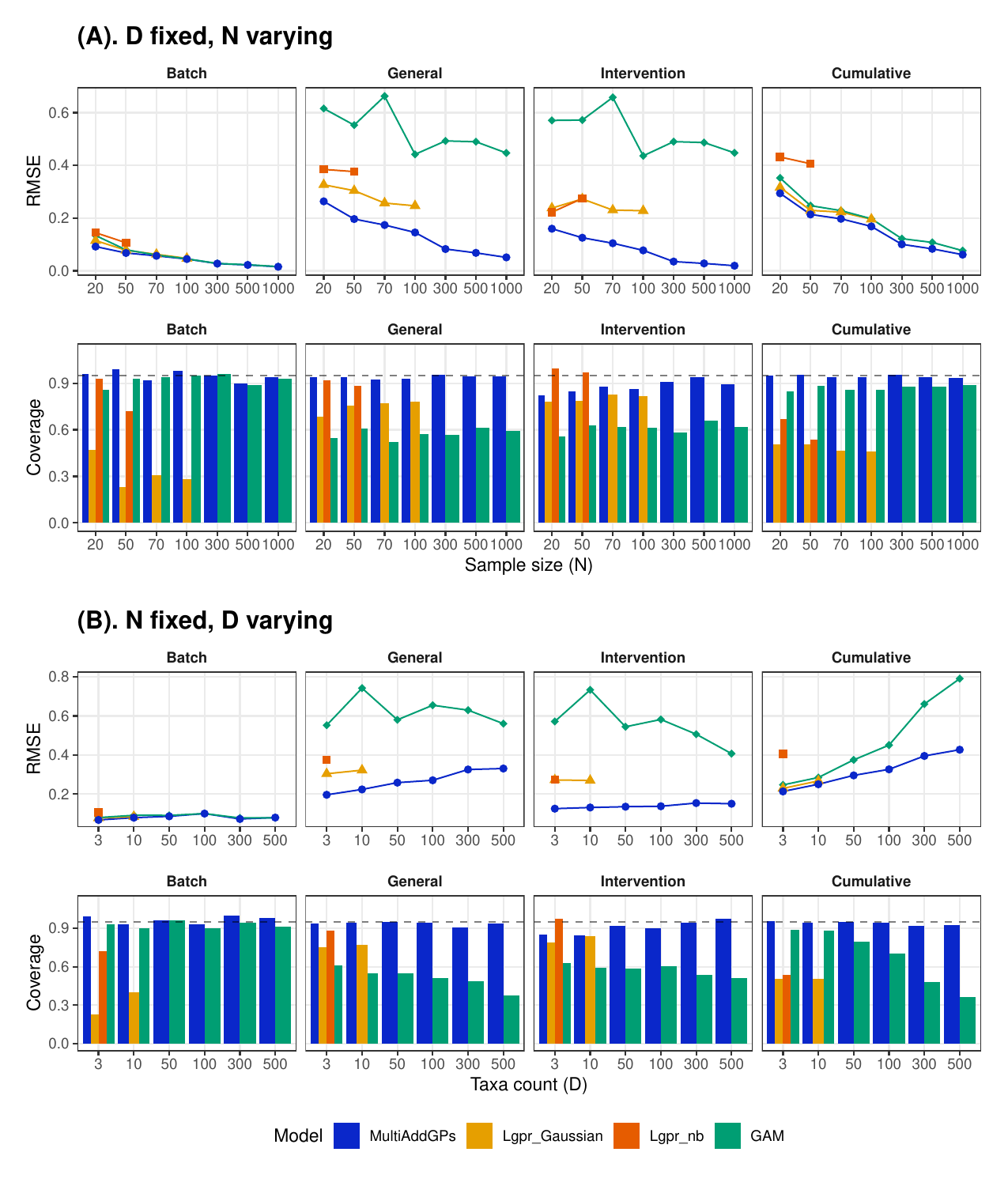}
\figurecaptionsheet
    \caption{\textbf{Simulation performance under varying sample size and taxa count.} Performance is assessed by varying sample size $N$ (Panel A, fixed $D=3$) 
    and taxa count $D$ (Panel B, fixed $N=50$). Top and bottom rows show RMSE 
    and 95\% interval coverage, respectively, based on 100 replicates per setting. 
    While MultiAddGPs and GAM covered all tested $(D, N)$ configurations, Lgpr was restricted 
    to smaller configurations due to its steep computational scaling.}
    \label{m-fig:full-metric}
    \par\smallskip
    {\def\baselinestretch{1.8}\normalsize\raggedright\noindent \textbf{Alt text:} Line plots show estimation error, and bar plots show interval coverage for batch, general, intervention, and cumulative effects. Panel A varies sample size; Panel B varies taxa count. MultiAddGPs generally has decreasing error as sample size increases and maintains coverage near the 95\% target. GAM shows lower coverage, particularly for the general and intervention effects. Lgpr results cover only smaller configurations.\par}
\end{figure}

Figure~\ref{m-fig:full-metric} summarizes estimation accuracy (RMSE) and uncertainty calibration (empirical coverage) for each additive component and their sum.
As \(N\) increases, MultiAddGPs exhibits decreasing RMSE and near-nominal coverage, indicating well-calibrated uncertainty.
GAM shows inflated RMSE and systematic under-coverage for nonlinear components.
Both Lgpr variants reduced nonlinear-component RMSE relative to GAM but showed greater undercoverage for the cumulative effect.
Increasing the number of taxa \(D\) yields only modest performance changes, reflecting the per-taxon structure of GAM and Lgpr. To assess sensitivity to GP hyperparameter specification, we additionally perturbed the magnitude (\(\sigma\)) and length-scale (\(\rho\)) parameters of the general and intervention components one at a time by \(\pm10\%\) for the \(D=3,\ N=50\) setting. Across 10 replicate datasets, RMSE and empirical coverage for the batch, general temporal, intervention, and cumulative effects remained similar to the baseline specification under all perturbations (see Figures 1 and 2 in~\ref{s-subsec:sim-lgpr-gam}). These results suggest that estimation accuracy and coverage are robust to moderate changes in the GP hyperparameters in this simulation setting.

\begin{figure}
    \centering
    \includegraphics[width=15cm]{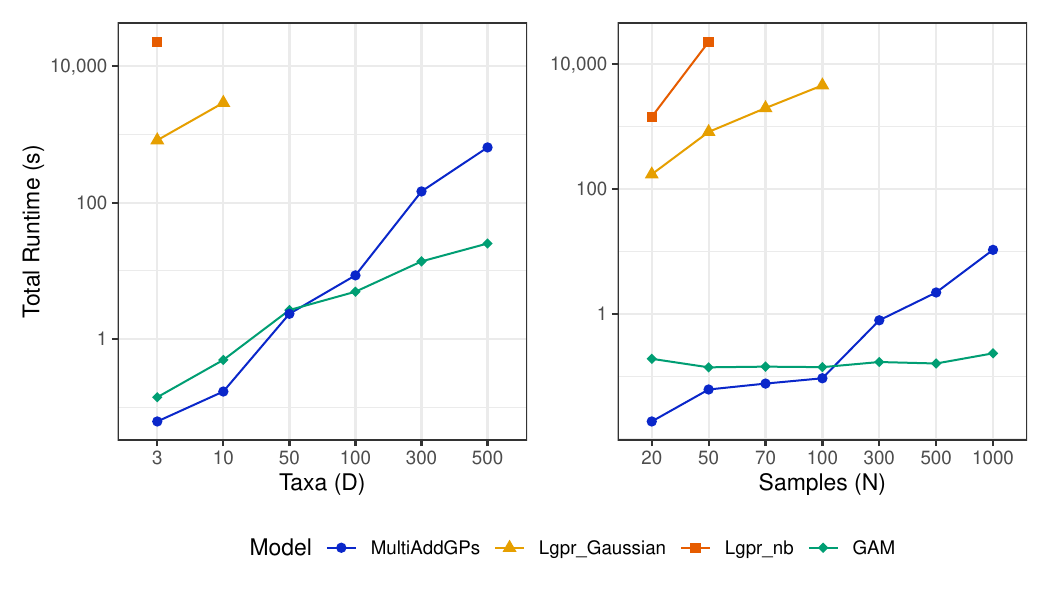 }
    \caption{\textbf{Total runtime of MultiAddGPs and competing methods.} The sample size is fixed at 50 in the left plot; the number of taxa is fixed at 3 in the right plot. Lgpr is shown only for configurations that completed within the computational budget, whereas MultiAddGPs and GAM were evaluated in all tested $(D, N)$ configurations.}
    \label{m-fig:runtime}
    \par\smallskip
    {\def\baselinestretch{1.8}\normalsize\raggedright\noindent \textbf{Alt text:} Two line plots show runtime in seconds on logarithmic vertical axes, varying taxa count on the left and sample size on the right. Both Lgpr variants require substantially more time than MultiAddGPs and GAM in the configurations evaluated. MultiAddGPs runtime increases with taxa count and sample size. GAM runtime increases with taxa count but remains relatively flat as sample size increases.\par}
\end{figure}

Figure~\ref{m-fig:runtime} reports mean estimated fitting time
as a function of the number of taxa $D$ (left) and sample size $N$ (right).
Runtime was defined differently for Bayesian and frequentist procedures.
For Bayesian methods (Lgpr and MultiAddGPs), we recorded elapsed model-fitting
time as time to obtain 2,000 posterior samples. For GAM, we recorded
the time required to fit the model and obtain parameter estimates.
The runtime calculations are detailed in~\ref{s-subsec:sim-lgpr-gam}.
The GAM baseline is the most computationally efficient method, as it
does not require posterior sampling. In contrast, both Lgpr variants
incur substantially higher computational cost and exhibit poor
scalability with increasing $D$ and $N$. MultiAddGPs demonstrates
markedly improved scalability relative to Lgpr, with runtime increasing
moderately as $D$ and $N$ grow.

To highlight the importance of count uncertainty, and to control for other potential differences between the above methods, we conducted an additional small-scale simulation study. We introduced an additional benchmark method, \textit{Normal Additive GPs} (NAddGPs), which are identical to MultiAddGPs but assumes Gaussian likelihood after a log-ratio transformation. Our result shows that the MultiAddGPs correctly recovered the true decomposition, whereas the NAddGPs substantially underestimated the amplitude of the nonlinear trend. Full details and results are provided in \ref{s-subsec:sim-Naddgps}.

\section{Application to Two Longitudinal Microbiome Data}
\label{m-sec:longitudinal}

We applied MultiAddGPs to two longitudinal gut microbiome studies to separate smooth background temporal variation from localized perturbation-associated dynamics. 

\subsection{Travel-associated perturbation and diarrheal illness in a human subject.}
\label{m-subsec:application-human}

We reanalyzed the daily gut microbiome time series from \citet{m-david2014host}, focusing on Subject A, who underwent prolonged international travel and experienced a diarrheal illness during that interval. This dataset is widely used to study within-person temporal microbiome variation~\citep{m-david2014host,m-gibbons2017two}, but prior analyses have characterized the travel interval as a broad community disturbance without explicitly separating travel-associated effects from background temporal variation. Using MultiAddGPs, we decomposed the trajectory into a smooth general temporal component and a travel-associated effect.

After preprocessing, the dataset contained
$D=47$ bacterial families observed across $N=329$ samples. We modeled the
latent trajectory as $\mathbf{F}_{\cdot n} = \mathbf{b}_0 + \mathbf{f}^{(\mathrm{general})}(t_n) + \mathbf{f}^{(\mathrm{travel})}(t_n)$,
where $\mathbf{f}^{(\mathrm{general})}$ represents smooth background temporal variation and
$\mathbf{f}^{(\mathrm{travel})}$ represents a localized travel-associated deviation.
Both components were assigned squared-exponential kernels. These GP kernels model temporal correlation between repeated observations within the subject. The general component was
modeled directly as a function of time, whereas the travel component used
a transformed time variable that concentrates its changes around the travel period.
To distinguish the two temporal processes, we constrained the travel-associated kernel
to vary more smoothly during the travel period and to have a smaller amplitude
parameter than the general-component kernel. Kernel hyperparameters were
estimated using the penalized marginal-likelihood procedure described in Section~\ref{m-subsec:method-hyperparameters}. Full prior and preprocessing specifications are provided in~\ref{s-sec:applications}.

\begin{figure}[!t]
     \centering
    \includegraphics[width=\linewidth]{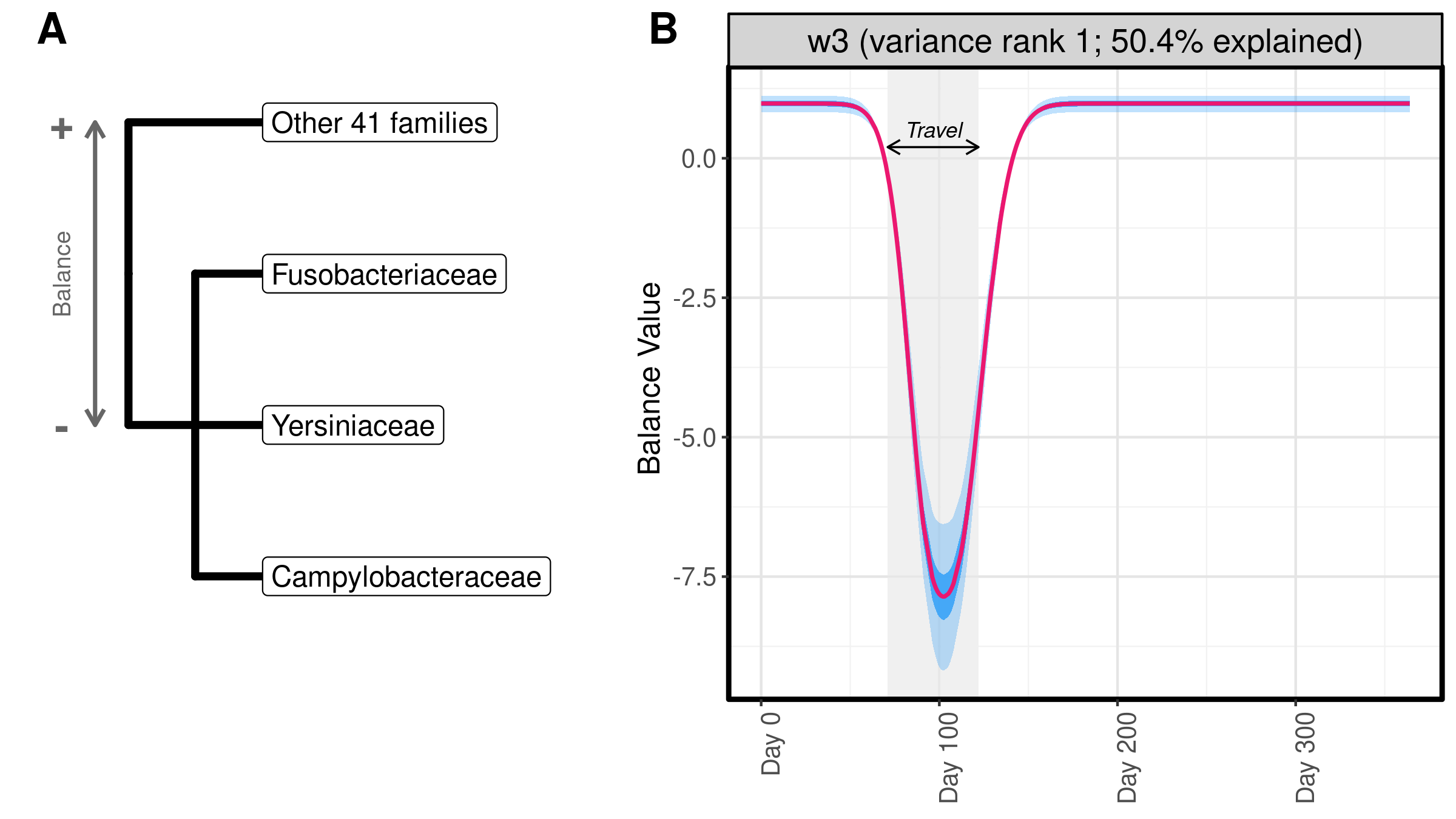}
    \caption{\textbf{Principal balance analysis of the isolated travel effect }($\mathbf{F}^{(\mathrm{travel})}$). (A) Taxon partition for the dominant balance, $w_3$, representing the scaled log-ratio of the numerator group ($+$) to the denominator group ($-$). (B) Posterior trajectory during international travel (shaded region), with posterior mean shown in pink and 50\% and 95\% credible intervals shown in dark and light blue, respectively. Balance $w_3$ reveals a sharp shift toward families containing taxa linked to inflammation and diarrhea (\textit{Fusobacteriaceae, Yersiniaceae, Campylobacteraceae}) relative to the 41-family comparison group.}    \label{m-fig:lifestyle_data_principalbalance}
    \par\smallskip
    {\def\baselinestretch{1.8}\normalsize\raggedright\noindent \textbf{Alt text:} Panel A contrasts 41 bacterial families on the positive side of balance w3 with Fusobacteriaceae, Yersiniaceae, and Campylobacteraceae on the negative side. Panel B shows the balance dropping sharply during the shaded travel period, reaching its minimum near day 100, and returning to its earlier level afterward. Shaded bands show posterior uncertainty. The balance explains 50.4\% of travel-associated variation.\par}
\end{figure}

To summarize the multivariate travel-associated effect, we constructed a dynamic principal balance analysis~\citep{m-silverman2018dynamic} 
from posterior draws of $\mathbf{F}^{(\mathrm{travel})}$. Specifically, for each posterior draw of the isolated travel component, we
computed taxon-to-taxon variation across time~\citep{m-aitchison1982statistical} and applied Ward clustering to obtain a tree.
We then formed a majority-rule consensus tree across draws and converted its sequential binary partition into an orthonormal isometric
log-ratio (ILR) basis, following the PhILR procedure of \citet{m-silverman2017phylogenetic}. Each ILR coordinate, or balance, is a scaled log
 ratio between the geometric means of the taxa in the two groups defined by a tree node. We ranked the balances according to their posterior
mean temporal variance and report the highest-ranked balance in Figure~\ref{m-fig:lifestyle_data_principalbalance}.

The dominant balance, $w_3$, explained 50.4\% (95\% credible interval 41.7\%--58.7\%) of the
posterior temporal variation attributable to the travel component and captured relative enrichment of \textit{Fusobacteriaceae}, \textit{Yersiniaceae}, and
\textit{Campylobacteraceae} as a group, alongside reciprocal depletion of the 41-family comparison group during travel.
These three families contain taxa with documented links to intestinal inflammation and diarrhea: invasive
\textit{Fusobacterium nucleatum} strains stimulate inflammatory responses in experimental
models~\citep{m-dharmani2011fusobacterium}, while \textit{Yersinia enterocolitica} and \textit{Campylobacter jejuni}
are established causes of diarrheal gastroenteritis~\citep{m-black1988experimental,m-marks1980yersinia}.
Taken together, this transient shift is consistent with the diarrheal disturbance reported during travel by
\citet{m-david2014host}, with the implicated families providing a plausible link to intestinal inflammation.

We assessed hyperparameter sensitivity using eight one-at-a-time
$\pm10\%$ perturbations of the travel and general GP amplitudes ($\sigma$) and
length scales ($\rho$). Holding the reference direction of $w_3$ fixed, its mean
share of travel-associated temporal variation ranged from 50.2\% to 52.1\%,
compared with 50.4\% at baseline. Posterior means and 95\% credible intervals
are reported in~\ref{s-sec:applications}.

\subsection{Starvation perturbation and recovery in an artificial gut.}
\label{m-subsec:application-artificial}

We next applied MultiAddGPs to a longitudinal artificial gut experiment with four physically isolated vessels~\citep{m-silverman2018dynamic}, each inoculated with an identical human fecal sample. The study includes 537 samples collected at irregular time points. A feed-starvation intervention was applied to Vessels 1 and 2 during Days 11--12, while Vessels 3 and 4 served as non-starved controls. Although prior analyses documented prolonged post-starvation recovery, they did not statistically separate starvation-induced disruption from baseline temporal drift.

After preprocessing, the analysis included $D=10$ bacterial families and
$N=537$ observations across the four vessels. We represented the four vessel
time series jointly within a single MultiAddGP model. Each vessel $v$ was assigned
a vessel-specific smooth baseline trajectory, $\mathbf{f}^{(\mathrm{base},v)}(t)$,
and the two starved vessels $v\in\{1,2\}$ additionally included a vessel-specific
starvation-associated component, $\mathbf{f}^{(\mathrm{disrupt},v)}(t)$. Thus, for
an observation $n$ from vessel $v_n$, we have
$ \mathbf{F}_{\cdot n} = \sum_{v=1}^{4} I(v_n=v) \left[\mathbf{f}^{(\mathrm{base},v)}(t_n) + I(v\in\{1,2\}) \mathbf{f}^{(\mathrm{disrupt},v)}(t_n) \right].$ Because the vessels were physically isolated, vessel-specific components were
modeled with block-diagonal covariance across vessels, so observations from
different vessels had zero temporal covariance while all vessel trajectories
were represented jointly within the same latent matrix.

Baseline trajectories were modeled using squared-exponential kernels, whereas
the starvation-associated components were modeled using rational-quadratic
kernels gated to zero before Day~11, with a $10^{-10}$ diagonal addition for numerical stability (implementation details in~\ref{s-subsec:artificial_data}). The baseline process was constrained to vary more smoothly and with
smaller magnitude than the starvation-associated process, allowing the latter
to capture localized post-starvation dynamics. Hyperparameters were estimated
using penalized marginal likelihood, with full prior specifications provided
in~\ref{s-sec:applications}.

Posterior summaries of the cumulative latent trajectory $\mathbf{F}$ (long-term trend plus starvation effect) and the isolated starvation component for \textit{Rikenellaceae} and \textit{Fusobacteriaceae} are shown in Figure~\ref{m-fig:artifical_decomposion} (full results for all taxa are provided in Supplement~\ref{s-sec:applications}). Consistent with prior reports, \textit{Rikenellaceae} exhibits a marked decline during the starvation window followed by gradual recovery toward baseline~\citep{m-silverman2018dynamic,m-sun2024predicting}. Decreases in \textit{Rikenellaceae} have also been reported in obesity studies~\citetext{\citealp*{m-okeke2014role}; \citealp{m-peters2018taxonomic,m-tavella2021elevated}}. The parallel patterns observed here raise the possibility that host-mediated mechanisms—absent in this \textit{in vitro} system—may contribute to obesity-associated shifts.

Beyond reproducing known patterns, the additive decomposition reveals additional structure. In particular, the posterior starvation component suggests an initial decline in \textit{Rikenellaceae} followed by an over-correction before returning toward pre-starvation levels. Such over-corrections are consistent with ecological predator–prey models that predict damped oscillatory responses to perturbations~\citep{m-samuelson1971generalized}. To our knowledge, this type of behavior has not previously been identified in \textit{in vitro} microbiome systems.

Across both studies, these results illustrate how additive decomposition within a multinomial logistic-normal framework enables separation of smooth background dynamics from localized perturbations.

\begin{figure}[!t]
     \centering
    \includegraphics[width=\linewidth]{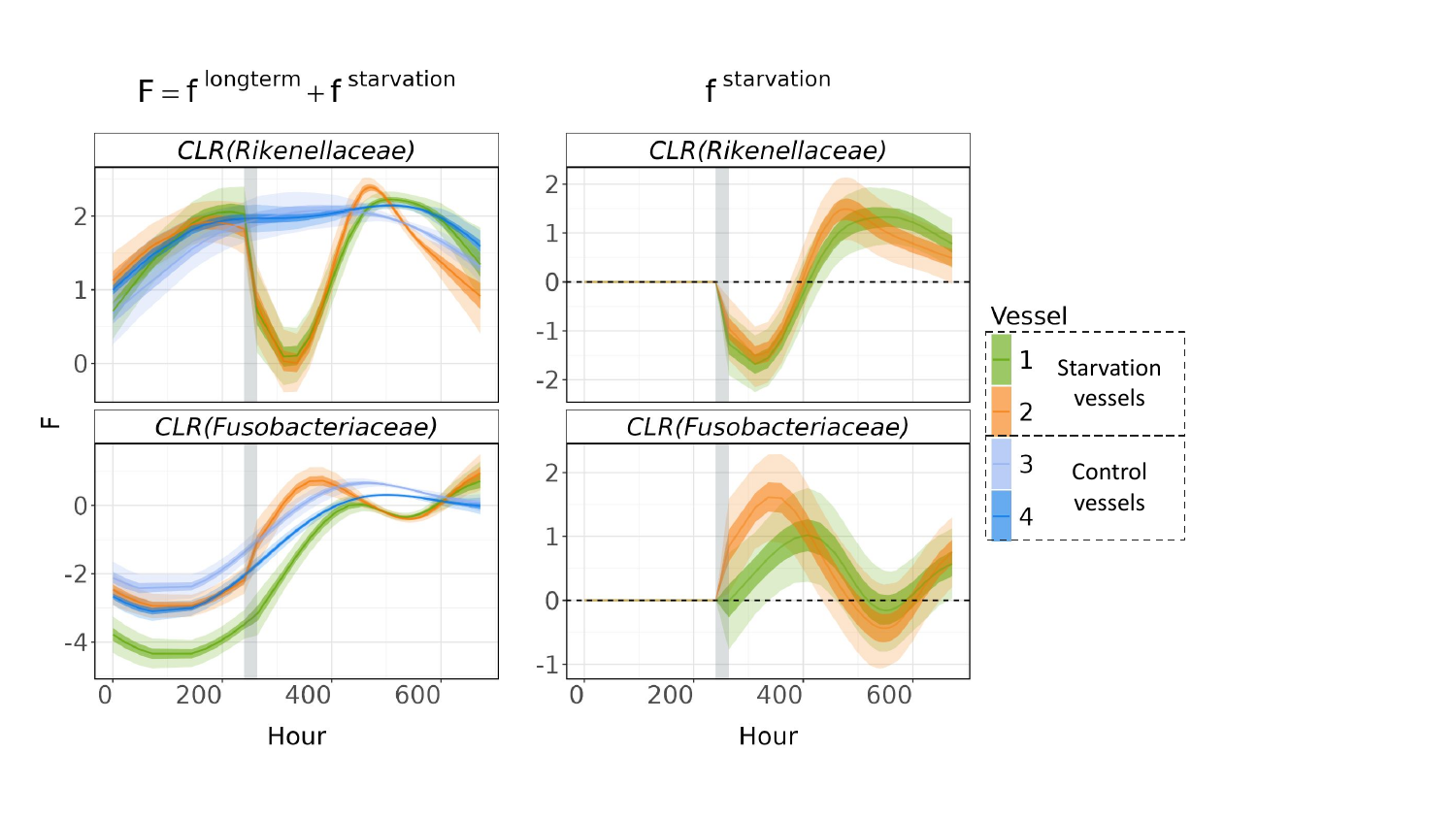}
    \caption{\textbf{MultiAddGPs posterior summaries for the smoothed latent state $\mathbf{F}$ and the starvation effect $\mathbf{F}^{(\mathrm{starvation})}$ in the families \textit{Rikenellaceae} and \textit{Fusobacteriaceae}.} As described in the main text, $\mathbf{F}$ is modeled as the sum of a slowly varying temporal trend and a short-lived, more variable starvation effect in Vessels 1 and 2 (beginning on Day 11). For each of the four artificial gut vessels, we show posterior means, 50\% and 95\% CI on the centered log-ratio (CLR) scale. The starvation period (Days 11--12) in Vessels 1 and 2 is shaded gray.} 
    \label{m-fig:artifical_decomposion}
    \par\smallskip
    {\def\baselinestretch{1.8}\normalsize\raggedright\noindent \textbf{Alt text:} Four plots show Rikenellaceae in the upper row and Fusobacteriaceae in the lower row, with time in hours along the horizontal axes. Left panels show combined temporal trajectories for four vessels; right panels isolate starvation effects in the two treated vessels. Following starvation, Rikenellaceae shows a negative effect followed by a positive rebound. Fusobacteriaceae initially shows a positive effect that subsequently declines. Shaded bands show posterior uncertainty.\par}
\end{figure}

\section{Discussion}
\label{m-sec:discussion}

Motivated by the need to separate overlapping sources of variation in longitudinal
microbiome data, we introduced MultiAddGPs, a semi-parametric Bayesian
multinomial logistic-normal framework that extends MLTP models to additive
combinations of linear and Gaussian process components. We first prove that our MultiAddGPs
remain within the MLTP class and derive the collapsed and uncollapsed representations needed
for scalable inference with the CU sampler. We then developed an efficient
back-sampling procedure to recover posterior draws of the individual additive
components, enabling direct inference on processes such as background temporal
variation and intervention-associated dynamics. We also developed approximate
maximum marginal likelihood and penalized maximum marginal likelihood procedures
for hyperparameter estimation in MLTP models. These procedures reuse quantities
already computed in the Laplace approximation and therefore add little
computational cost to posterior inference. Together, these developments provide
a scalable framework for decomposing multivariate compositional dynamics while
accounting for multinomial sampling uncertainty. Our methods are implemented in
the \textit{fido} R package on CRAN.

While our methods are inspired by challenges in microbiome data, MultiAddGPs could be applied broadly to other settings where count compositional data is collected, including gene expression studies~\citep{m-fernandes2014unifying}, macro-ecology~\citep*{m-billheimer2001statistical}, and even political polling~\citep{m-topirceanu2020polling}. Still, several key challenges, extensions, and future directions remain.

One possible modeling extension is to incorporate spline-based nonlinear effects within
the MLTP framework. Because previous MLTP models already accommodate linear regression
\citep{m-silverman2022bayesian}, spline basis functions could in principle be included in
the design matrix. However, flexible spline models generally require smoothing penalties,
such as roughness penalties based on integrated squared second derivatives
\citep{m-wood2017generalized}. Incorporating such penalties within the MLTP framework
requires an appropriate prior precision structure for the spline coefficients and is therefore
not achieved by simply adding spline basis terms. Gaussian process priors provide a natural
alternative in the present work because smoothness is encoded directly through the kernel
while staying within the MLTP framework.

A separate practical challenge concerns hyperparameter estimation. The approximate MML
and penalized MML procedures developed here provide a flexible approach for estimating
hyperparameters within kernel functions and, more generally, within other components of
an MLTP model, such as the log-ratio transformation. In practice, we have found Bayesian
optimization to be effective for small to moderate numbers of hyperparameters (e.g., fewer
than 10). However, further methodological development may be needed when many
hyperparameters must be estimated simultaneously or when the marginal-likelihood surface
is particularly difficult to optimize.

Besides the challenge of calculating MML and penalized MML estimates, alternative methods of selecting hyperparameters may be needed. In particular, some researchers may require fully Bayesian solutions that can quantify uncertainty in model hyperparameters. This is a notable advantage of the approach of \citet{m-cheng2019additivegp}. While their model ignores counting variation, and assumes the data is transformed Gaussian, their method provides fully Bayesian inference of hyperparameters via the slice sampler of \cite{m-murray2010slicegauss}. We expect that the same slice-sampling approach could be used for MultiAddGPs, replacing the Metropolis-Hastings Steps used in that method with the CU sampler with marginal Laplace approximation. 

A further modeling limitation concerns structural zeros.
MultiAddGPs accommodates observed zeros through multinomial sampling but does not explicitly distinguish structural (biological) zeros from sampling zeros. Its logistic-normal representation assigns positive latent relative abundances, so true absences can only be approximated by very low abundance; this limitation may affect inference for taxa with systematic biological absences~\citep{m-silverman2020naught}.

Finally, a computational limitation concerns the accuracy of the Laplace approximation in the CU sampler. In \cite{m-silverman2022bayesian}, we proved that the error rate of this Laplace approximation is of order \(O_{p}\left\{(D-1)\sum_{n=1}^{N}\lambda_n^{-1}\right\}\). This result implies that the approximation works well when the average number of counts per sample (e.g., the sequencing depth) is high. This is the case with modern sequencing technologies. In contrast, we expect the approximation will be poor when there are few counts per sample, for example, when observations are categorical (\(\sum_{d=1}^{D}Y_{dn}=1\)). In this scenario, alternative methods, such as the Debiased Multinomial Dirichlet Bootstrap from~\citet*{m-saxena2024mlndlm}, may be necessary.

\backmatter

\section*{Acknowledgments}
We thank Rachel Silverman, Hangzhi Guo, and an anonymous referee for helpful comments. This research was supported by NIGMS and NIDDK of the National Institutes of Health (NIH) (R01GM148972 and R01DK116187), with 50\% federal and 50\% non-governmental funding. The content is solely the responsibility of the authors and does not necessarily represent the official views of the NIH. The authors used OpenAI’s Codex for code debugging and manuscript editing. \textbf{Conflict of Interest.} The authors declare no conflicts of interest. \vspace*{-8pt}

\section*{Supplementary Materials}
Web Appendices and Figures, analysis code, and processed artificial-gut data
are available with this paper at the \textit{Biometrics} website on Oxford
Academic. The code and data are also available at
\href{https://github.com/Silverman-Lab/MultiAddGPs}{\texttt{MultiAddGPs}};
MultiAddGPs is implemented in \texttt{fido}
(\href{https://CRAN.R-project.org/package=fido}{CRAN}).

\section*{Data Availability}
The human longitudinal microbiome sequence data are available through the
European Nucleotide Archive (ERP006059), and Subject A's nutritional metadata
under SAMEG179160. Artificial-gut sequence data are available through the NCBI
Sequence Read Archive (accession SRP140877 and BioProject PRJNA450809). Code and processed
artificial-gut data are available as described under Supplementary Materials.

\label{m-lastpage}

\processdelayedfloats
\clearpage
\makeatletter
\global\backmatterfalse
\makeatother
\raggedbottom
\setcounter{page}{1}
\setcounter{section}{0}
\setcounter{subsection}{0}
\setcounter{equation}{0}
\setcounter{figure}{0}
\setcounter{postfigure}{0}
\setcounter{posttable}{0}
\setcounter{table}{0}
\setcounter{algorithm}{0}
\setcounter{theorem}{0}
\setcounter{definition}{0}
\renewcommand{\thesection}{Web Appendix \Alph{section}}
\renewcommand{\thefigure}{\arabic{figure}}
\renewcommand{\thetable}{\arabic{table}}
\renewcommand{\figurename}{Web Figure}
\renewcommand{\tablename}{Web Table}
\renewcommand{\theHsection}{supp.\arabic{section}}
\renewcommand{\theHequation}{supp.\arabic{equation}}
\renewcommand{\theHfigure}{supp.\arabic{figure}}
\renewcommand{\theHtable}{supp.\arabic{table}}
\providecommand{\theHalgorithm}{supp.\arabic{algorithm}}
\makeatletter
\renewcommand{\fnum@figure}{\figurename~\thefigure}
\renewcommand{\fnum@table}{\tablename~\thetable}
\makeatother
\renewenvironment{proof}[1][Proof]{\par\noindent\textit{#1.}\ }{\hfill$\square$\par}
\let\mathbb\arxivMathbb
\renewcommand{\R}{\mathbb{R}}
\renewcommand{\S}{\mathbb{S}}
\renewcommand{\Z}{\mathtt{Z}}
\let\inf\arxivSuppInf
\let\sup\arxivSuppSup

\begin{center}
\large\textbf{Supplementary Materials for Bayesian Multinomial Logistic
Normal Additive Regression for Microbiome Data Analysis by Tinghua Chen,
Michelle Pistner Nixon, Lawrence A. David, and Justin D. Silverman.}
\end{center}

\title[Supplementary Materials: Additive Gaussian Processes]{Bayesian Multinomial Logistic Normal Additive Regression for Microbiome Data Analysis}

\section{Review of the Collapse-Uncollapse (CU) Sampler}
\label{s-sec: Marginally-LTP}
Just as Gaussian processes can be defined based on the marginal properties of the multivariate normal, matrix normal processes and matrix-t processes can be defined by the marginal properties of the matrix normal and matrix-t distributions~\citep{s-silverman2022bayesian}. Matrix-t processes generalize Student-t processes and Gaussian processes~\citep{s-silverman2022bayesian}. 

\begin{definition}[Matrix-T Process]
  A stochastic process \(\mathbf{Y} \sim TP(\nu, \mathbf{M}, \mathbf{V}, \mathbf{A})\) defined on the set \(\mathcal{W} = \mathcal{W}^{(1)} \times \mathcal{W}^{(2)}\) is a matrix-T process if \(\mathbf{Y}\) evaluated on any two finite subsets \(\mathcal{X}^{(1)} \subset \mathcal{W}^{(1)}\) and \(\mathcal{X}^{(2)} \subset  \mathcal{W}^{(2)}\) is a random matrix of dimension \(|\mathcal{X}^{(1)}| \times |\mathcal{X}^{(2)}|\) that follows a matrix-T distribution: \(\mathbf{Y} \sim T(\nu, \mathbf{M, V, A})\). \(\nu\) is a scalar value strictly greater than zero.
  Let \(x_{i}^{(1)}, x_{j}^{(1)} \in \mathcal{X}^{(1)}\) and \(x_{i}^{(2)}, x_{j}^{(2)} \in \mathcal{X}^{(2)}\). \(\mathbf{M}_{ij}=\mathbf{M}(x_{i}^{(1)} , x_{j}^{(2)})\) is the location function, and \(\mathbf{V}_{ij}=\mathbf{V}(x_{i}^{(1)}, x_{j}^{(1)})\) and \(\mathbf{A}_{ij}=\mathbf{A}(x_{i}^{(2)}, x_{j}^{(2)})\) are row and column scale kernels, respectively.
\end{definition}

Latent Matrix-T Processes (LTP) generalize Matrix-T processes.
\begin{definition}[Latent Matrix-T Processes]
\(\mathbf{Y}\) is said to be an LTP if
\begin{align*}
\mathbf{Y} &\sim g(\boldsymbol{\Pi}), \\
  \boldsymbol{\Pi}&=\phi^{-1}(\mathbf{H}), \\
  \mathbf{H} &\sim TP(\nu, \mathbf{M}, \mathbf{V}, \mathbf{A}), 
\end{align*}
where \(g\) is any distribution depending on parameters \(\boldsymbol{\Pi}\) and \(\phi\) is a known transform. 
\end{definition}

A stochastic process \(\mathbf{Y}\) is Marginally LTP (MLTP) if there exists a joint distribution \(p(\mathbf{Y}, \mathbf{H}, \mathbf{\Phi})\) such that the marginal \(p(\mathbf{Y}, \mathbf{H})\) is a LTP. This definition of MLTPs is key to efficient inference. We call the marginal \(p(\mathbf{Y}, \mathbf{H})\) the \textit{collapsed form} and \(p(\mathbf{\Phi} \mid \mathbf{Y}, \mathbf{H})\) the \textit{uncollapsed form}. The posterior of an MLTP factors as
\[p(\mathbf{H},\mathbf{\Phi} \mid \mathbf{Y})=p(\mathbf{\Phi}\mid \mathbf{H}, \mathbf{Y})p(\mathbf{H} \mid \mathbf{Y}),\]
with the uncollapsed form as the first term and the posterior of the collapsed form as the second. 
As the collapsed form is rarely conjugate, techniques such as MCMC can be used to obtain samples from its posterior. Then, conditioned on those samples, the uncollapsed form can be used to obtain samples from the joint posterior. Especially when \(\boldsymbol{\Phi}\) is high-dimensional, this \textit{Collapse-Uncollapse} sampler can be much more efficient than common alternatives~\citep{s-silverman2022bayesian}. Still, the most substantial enhancements occur when approximations to the collapsed form are considered. 
We have developed a Laplace approximation for the collapsed form of MLTPs:
\begin{equation}
    p(\text{vec}(\mathbf{H})\mid \mathbf{Y})\approx \mathcal{N}(\text{vec}(\widehat{\mathbf{H}}),\nabla^{-2}[\text{vec}(\widehat{\mathbf{H}})]),
\end{equation}
where \(\widehat{\mathbf{H}}\) denotes the \textit{maximum a posteriori (MAP)} estimate of the collapsed form and \(\nabla^{-2}[\text{vec}(\widehat{\mathbf{H}})]\) denotes the inverse Hessian of the collapsed form evaluated at the MAP estimate. The analytical expressions for the gradient and Hessian of the collapsed form are provided in~\ref{s-sec:LA}. 
Focusing on applications to MultiAddGPs, we proved error bounds on the Laplace approximation and provided simulation and real analyses showing that the approximation was extremely accurate in the context of microbiome data analysis. 
Beyond the accuracy of posterior calculations, we showed that this CU sampler with the Laplace approximation (simply referred to as the CU sampler in the following text) was often 4-5 orders of magnitude faster than MCMC and 1-2 orders of magnitude faster than black-box variational inference while also being more accurate than the latter. 

\section{MultiAddGPs as Marginally Latent Matrix-t Process (MLTPs)}
\label{s-sec:proof}
\subsection{Distributional conventions}
\label{s-sec:distribution-conventions}
Let $P=D-1$ be the number of log-ratio coordinates. Throughout, $IW(\boldsymbol{\Xi},\zeta)$ denotes the inverse-Wishart distribution with scale $\boldsymbol{\Xi}>0$ and degrees of freedom $\zeta>P-1$, with density
\begin{equation}
p(\boldsymbol{\Sigma})=
\frac{|\boldsymbol{\Xi}|^{\zeta/2}}{2^{\zeta P/2}\Gamma_P(\zeta/2)}
|\boldsymbol{\Sigma}|^{-(\zeta+P+1)/2}
\exp\!\left\{-\tfrac12\operatorname{tr}(\boldsymbol{\Xi}\boldsymbol{\Sigma}^{-1})\right\},
\quad \boldsymbol{\Sigma}>0.
\label{s-eq:iw-density}
\end{equation}
Here $\Gamma_P$ is the multivariate gamma function. In particular, $\mathbb{E}(\boldsymbol{\Sigma})=\boldsymbol{\Xi}/(\zeta-P-1)$ when $\zeta>P+1$.
For a $P\times N$ matrix, $T(\nu,\mathbf{M},\mathbf{V},\mathbf{A})$ has degrees of freedom $\nu>0$, location $\mathbf{M}$, and positive-definite row and column scale matrices $\mathbf{V}$ and $\mathbf{A}$. Its density is
\begin{align}
p(\mathbf{H})
&=\frac{\Gamma_P\{(\nu+N+P-1)/2\}}{\pi^{PN/2}\Gamma_P\{(\nu+P-1)/2\}|\mathbf{V}|^{N/2}|\mathbf{A}|^{P/2}}
\nonumber\\
&\quad\times\left|\mathbf{I}_P+\mathbf{V}^{-1}(\mathbf{H}-\mathbf{M})
\mathbf{A}^{-1}(\mathbf{H}-\mathbf{M})^T\right|^{-(\nu+N+P-1)/2}.
\label{s-eq:matrix-t-density}
\end{align}
Thus, integrating $\boldsymbol{\Sigma}$ out of
$\mathbf{H}\mid\boldsymbol{\Sigma}\sim\mathcal{MN}(\mathbf{M},\boldsymbol{\Sigma},\mathbf{A})$ and
$\boldsymbol{\Sigma}\sim IW(\boldsymbol{\Xi},\zeta)$ gives
$\mathbf{H}\sim T(\zeta-P+1,\mathbf{M},\boldsymbol{\Xi},\mathbf{A})$.
The degrees of freedom therefore satisfy $\nu=\zeta-D+2$, and the determinant exponent is $-(\zeta+N)/2$.
The corresponding uncollapsed covariance posterior has inverse-Wishart degrees of freedom $\zeta+N$.

\subsection{Derivation of Collapsed form}

\begin{theorem}
  MultiAddGPs are MLTPs with parameters  $\boldsymbol{\Phi} = \{\mathbf{F},\mathbf{B}, \mathbf{F}^{(1)}, \dots, \mathbf{F}^{(K)}, \boldsymbol{\Sigma}\}$. The collapsed form \(p( \mathbf{Y},\mathbf{H})\) is an LTP:
  \begin{align*}
  \mathbf{Y}_{\cdot n} &\sim \operatorname{Multinomial}(\lambda_n, \boldsymbol{\Pi}_{\cdot n}), \\ 
    \boldsymbol{\Pi}_{\cdot n} &= \phi^{-1}(\mathbf{H}_{\cdot n}), \\
    \mathbf{H} &\sim TP(\nu, \mathbf{M}, \mathbf{V}, \mathbf{A}),
  \end{align*}
 where $\phi$ is an invertible log-ratio transformation, $  \mathbf{M}=\boldsymbol{\Theta}^{(0)}\mathbf{X} + \sum_{k=1}^K\boldsymbol{\Theta}^{(k)}(\mathbf{Z}^{(k)})$, $\mathbf{V} = \boldsymbol{\Xi}$, and $\mathbf{A} = \mathbf{X}^{T} \boldsymbol{\Gamma}^{(0)} \mathbf{X} + \sum_{k=1}^{K} \boldsymbol{\Gamma}^{(k)}(\mathbf{Z}^{(k)}) + \mathbf{I}_N, \nu =\zeta-D+2$.
 \end{theorem}

\begin{proof}
    By the definition of an MLTP~\citep{s-silverman2022bayesian}, it suffices to show that the marginal $p(\mathbf{H},\mathbf{Y})$ is an LTP. To begin, we note that the latent-state and additive mean models in Equations~\eqref{m-eqs:State_F} and~\eqref{m-eqs:F} of the main text (along with their priors) can alternatively be written as:
    \begin{align}
        \mathbf{H} &= \mathbf{F}+\mathbf{E}^{\mathbf{H}}, \hspace{1.2cm} \mathbf{E}^{\mathbf{H}} \sim \mathcal{MN}(0,\boldsymbol{\Sigma},\mathbf{I}_N), \label{s-eq:H}\\
        \mathbf{F} &= \mathbf{BX} + \sum_{k=1}^{K} \mathbf{F}^{(k)}, \label{s-eq:F}\\
        \mathbf{B} &= \boldsymbol{\Theta}^{(0)} + \mathbf{E}^{\mathbf{B}}, \hspace{0.8cm} \mathbf{E}^{\mathbf{B}} \sim \mathcal{MN}(0,\boldsymbol{\Sigma},\boldsymbol{\Gamma}^{(0)}), \label{s-eq:B}\\
        \mathbf{F}^{(k)} &= \boldsymbol{\Theta}^{(k)}(\mathbf{Z}^{(k)}) +  \mathbf{E}^{\mathbf{F}^{(k)}}, \hspace{0.3cm} \mathbf{E}^{\mathbf{F}^{(k)}} \sim \mathcal{MN}(0,\boldsymbol{\Sigma},\boldsymbol{\Gamma}^{(k)}(\mathbf{Z}^{(k)})), \label{s-eq:f}\\
        \boldsymbol{\Sigma} &\sim IW(\mathbf{\Xi}, \zeta). \label{s-eq:Sigma}
    \end{align}
    Assume that $\mathbf{E}^{\mathbf{B}},\mathbf{E}^{\mathbf{F}^{(1)}},\dots,\mathbf{E}^{\mathbf{F}^{(K)}}$ are mutually independent conditional on $\boldsymbol{\Sigma}$, and independent of $\mathbf{E}^{\mathbf{H}}$ conditional on $\boldsymbol{\Sigma}$. Using this form, in combination with the affine transformation property of the matrix normal distribution, we can rewrite the latent $\mathbf{F}$ in the following form:
    \begin{align}
        \mathbf{F} &=\boldsymbol{\Theta}^{(0)}\mathbf{X} + \sum_{k=1}^{K}\boldsymbol{\Theta}^{(k)}(\mathbf{Z}^{(k)}) +\mathbf{E}^{\mathbf{B}}\mathbf{X} + \sum_{k=1}^{K}\mathbf{E}^{\mathbf{F}^{(k)}}, \label{s-eq:Marginalize_B_f_1}\\
        \mathbf{F} &= \boldsymbol{\Theta}^{(0)}\mathbf{X} + \sum_{k=1}^{K}\boldsymbol{\Theta}^{(k)}(\mathbf{Z}^{(k)}) + \mathbf{E}^{\mathbf{F}} , \hspace{0.2cm} \mathbf{E}^{\mathbf{F}} \sim \mathcal{MN}(0, \boldsymbol{\Sigma},\mathbf{X}^T\boldsymbol{\Gamma}^{(0)}\mathbf{X}+\sum_{k=1}^{K}\boldsymbol{\Gamma}^{(k)}(\mathbf{Z}^{(k)})).  \label{s-eq:Marginalize_B_f_2}
    \end{align}
    Thus we may rewrite Equations (\ref{s-eq:H}) - (\ref{s-eq:Sigma}) as
    \begin{align}
        \mathbf{H} &= \mathbf{F}+\mathbf{E}^{\mathbf{H}}, \hspace{1.2cm} \mathbf{E}^{\mathbf{H}} \sim \mathcal{MN}(0,\boldsymbol{\Sigma},\mathbf{I}_N), \label{s-eq:H2}\\
       \mathbf{F} &= \boldsymbol{\Theta}^{(0)}\mathbf{X} + \sum_{k=1}^{K}\boldsymbol{\Theta}^{(k)}(\mathbf{Z}^{(k)}) + \mathbf{E}^{\mathbf{F}} , \hspace{0.2cm} \mathbf{E}^{\mathbf{F}} \sim \mathcal{MN}(0, \boldsymbol{\Sigma},\mathbf{X}^T\boldsymbol{\Gamma}^{(0)}\mathbf{X}+\sum_{k=1}^{K}\boldsymbol{\Gamma}^{(k)}(\mathbf{Z}^{(k)})),   \label{s-eq:F2} \\
        \boldsymbol{\Sigma} &\sim IW(\boldsymbol{\Xi}, \zeta). \label{s-eq:Sigma2}
    \end{align}
    Following the result from \cite{s-gupta2018matrix}, we can marginalize over $\mathbf{F}$ and $\boldsymbol{\Sigma}$ in Equations (\ref{s-eq:F2}) and (\ref{s-eq:Sigma2}) to get 
    \begin{align*}
        \mathbf{H} \sim TP(\zeta-D+2,\boldsymbol{\Theta}^{(0)}\mathbf{X} + \sum_{k=1}^{K}\boldsymbol{\Theta}^{(k)}(\mathbf{Z}^{(k)}) ,\mathbf{\Xi},\mathbf{X}^T\boldsymbol{\Gamma}^{(0)}\mathbf{X}+\sum_{k=1}^{K}\boldsymbol{\Gamma}^{(k)}(\mathbf{Z}^{(k)}) + \mathbf{I}_N).
    \end{align*}
    Finally, incorporating the multinomial observation model and log-ratio transformation in Equations~\eqref{m-eqs:multinomial} and~\eqref{m-eqs:logratiotransform} of the main text allows us to write the marginalized form of the MultiAddGP model as an LTP
    \begin{align*}
          \boldsymbol{Y} &\sim g(\boldsymbol{\Pi}, \lambda), \\
          \boldsymbol{\Pi}&=\phi^{-1}(\boldsymbol{H}), \\
          \boldsymbol{H} &\sim TP(\nu, \boldsymbol{M}, \boldsymbol{V}, \boldsymbol{A}), 
    \end{align*}
    where $g$ is product multinomial, $\phi$ is an invertible log-ratio transformation, $  \mathbf{M}=\boldsymbol{\Theta}^{(0)}\mathbf{X} + \sum_{k=1}^K\boldsymbol{\Theta}^{(k)}(\mathbf{Z}^{(k)})$, $\mathbf{V} = \boldsymbol{\Xi}$, and $\mathbf{A} = \mathbf{X}^{T} \boldsymbol{\Gamma}^{(0)} \mathbf{X} + \sum_{k=1}^{K} \boldsymbol{\Gamma}^{(k)}(\mathbf{Z}^{(k)}) + \mathbf{I}_N, \nu =\zeta-D+2$.
\end{proof} 

\subsection{Derivation of Uncollapsed form - Sample from $p(\mathbf{F}, \boldsymbol{\Sigma}|\mathbf{H}, \mathbf{Y})$}
\label{s-subsec:sample-F-Sigma}

Here, we demonstrate how to efficiently sample from the conditional posterior
\(p(\mathbf{F},\boldsymbol{\Sigma}\mid\mathbf{H},\mathbf{Y},
\mathbf{X},\mathbf{Z})\). Since \(\mathbf{F}\) and
\(\boldsymbol{\Sigma}\) are conditionally independent of \(\mathbf{Y}\)
given \(\mathbf{H}\), we have
\begin{align}
p(\mathbf{F},\boldsymbol{\Sigma}
\mid\mathbf{H},\mathbf{Y},\mathbf{X},\mathbf{Z})
&=
p(\mathbf{F},\boldsymbol{\Sigma}
\mid\mathbf{H},\mathbf{X},\mathbf{Z}) \nonumber\\
&=
p(\mathbf{F}\mid
\boldsymbol{\Sigma},\mathbf{H},\mathbf{X},\mathbf{Z})
p(\boldsymbol{\Sigma}\mid
\mathbf{H},\mathbf{X},\mathbf{Z}).
\end{align}

From the representation in Section~\ref{s-sec: Marginally-LTP}, define
\begin{align*}
\mathbf{M}
&=
\boldsymbol{\Theta}^{(0)}\mathbf{X}
+
\sum_{k=1}^{K}
\boldsymbol{\Theta}^{(k)}(\mathbf{Z}^{(k)}),\\
\boldsymbol{\Gamma}_{F}
&=
\mathbf{X}^{T}\boldsymbol{\Gamma}^{(0)}\mathbf{X}
+
\sum_{k=1}^{K}
\boldsymbol{\Gamma}^{(k)}(\mathbf{Z}^{(k)}).
\end{align*}
Using matrix-normal/inverse-Wishart conjugacy, define
\begin{align*}
\boldsymbol{\Gamma}_{N}
&=
\left[
\mathbf{I}_{N}
+
\boldsymbol{\Gamma}_{F}^{-1}
\right]^{-1},\\
\mathbf{F}_{N}
&=
\left[
\mathbf{H}
+
\mathbf{M}\boldsymbol{\Gamma}_{F}^{-1}
\right]
\boldsymbol{\Gamma}_{N},\\
\zeta_{N}
&=
\zeta+N,\\
\boldsymbol{\Xi}_{N}
&=
\boldsymbol{\Xi}
+
(\mathbf{H}-\mathbf{F}_{N})
(\mathbf{H}-\mathbf{F}_{N})^{T}\\
&\quad+
(\mathbf{F}_{N}-\mathbf{M})
\boldsymbol{\Gamma}_{F}^{-1}
(\mathbf{F}_{N}-\mathbf{M})^{T}.
\end{align*}
The conditional posterior distributions are therefore
\begin{align}
\boldsymbol{\Sigma}\mid
\mathbf{H},\mathbf{X},\mathbf{Z}
&\sim
\mathcal{IW}
\left(
\boldsymbol{\Xi}_{N},
\zeta_{N}
\right),
\label{s-eq:uncollapse-sigma}\\
\mathbf{F}\mid
\boldsymbol{\Sigma},\mathbf{H},\mathbf{X},\mathbf{Z}
&\sim
\mathcal{MN}
\left(
\mathbf{F}_{N},
\boldsymbol{\Sigma},
\boldsymbol{\Gamma}_{N}
\right).
\label{s-eq:uncollapse-F}
\end{align}
These updates provide the first uncollapse step of the CU sampler.

For completeness, the same Gaussian process representation can also be
used for prediction at unobserved input locations. Let \(o\) and \(u\)
index observed and unobserved sample locations, respectively, and write
\[
\mathbf{X}
=
\begin{bmatrix}
\mathbf{X}^{o} & \mathbf{X}^{u}
\end{bmatrix},
\qquad
\mathbf{Z}^{(k)}
=
\begin{bmatrix}
\mathbf{Z}^{(k),o} & \mathbf{Z}^{(k),u}
\end{bmatrix},
\qquad
\mathbf{F}
=
\begin{bmatrix}
\mathbf{F}^{o} & \mathbf{F}^{u}
\end{bmatrix}.
\]
The corresponding mean functions are
\begin{align*}
\mathbf{M}^{o}
&=
\boldsymbol{\Theta}^{(0)}\mathbf{X}^{o}
+
\sum_{k=1}^{K}
\boldsymbol{\Theta}^{(k)}
(\mathbf{Z}^{(k),o}),\\
\mathbf{M}^{u}
&=
\boldsymbol{\Theta}^{(0)}\mathbf{X}^{u}
+
\sum_{k=1}^{K}
\boldsymbol{\Theta}^{(k)}
(\mathbf{Z}^{(k),u}),
\end{align*}
and the column covariance is partitioned as
\[
\boldsymbol{\Gamma}_{F}
=
\begin{bmatrix}
\boldsymbol{\Gamma}_{F}^{oo}
&
\boldsymbol{\Gamma}_{F}^{ou}\\
\boldsymbol{\Gamma}_{F}^{uo}
&
\boldsymbol{\Gamma}_{F}^{uu}
\end{bmatrix},
\]
where
\begin{align*}
\boldsymbol{\Gamma}_{F}^{oo}
&=
(\mathbf{X}^{o})^{T}
\boldsymbol{\Gamma}^{(0)}
\mathbf{X}^{o}
+
\sum_{k=1}^{K}
\boldsymbol{\Gamma}^{(k)}
(\mathbf{Z}^{(k),o},\mathbf{Z}^{(k),o}),\\
\boldsymbol{\Gamma}_{F}^{uu}
&=
(\mathbf{X}^{u})^{T}
\boldsymbol{\Gamma}^{(0)}
\mathbf{X}^{u}
+
\sum_{k=1}^{K}
\boldsymbol{\Gamma}^{(k)}
(\mathbf{Z}^{(k),u},\mathbf{Z}^{(k),u}),\\
\boldsymbol{\Gamma}_{F}^{ou}
&=
(\mathbf{X}^{o})^{T}
\boldsymbol{\Gamma}^{(0)}
\mathbf{X}^{u}
+
\sum_{k=1}^{K}
\boldsymbol{\Gamma}^{(k)}
(\mathbf{Z}^{(k),o},\mathbf{Z}^{(k),u}),\\
\boldsymbol{\Gamma}_{F}^{uo}
&=
(\boldsymbol{\Gamma}_{F}^{ou})^{T}.
\end{align*}

Conditional on samples of \(\mathbf{F}^{o}\) and
\(\boldsymbol{\Sigma}\), prediction at the unobserved locations follows
from the standard conditional matrix-normal distribution:
\begin{align}
\mathbf{F}^{u}
\mid
\mathbf{F}^{o},\boldsymbol{\Sigma}
\sim
\mathcal{MN}
\Big(
&\mathbf{M}^{u}
+
(\mathbf{F}^{o}-\mathbf{M}^{o})
(\boldsymbol{\Gamma}_{F}^{oo})^{-1}
\boldsymbol{\Gamma}_{F}^{ou},
\boldsymbol{\Sigma}, \nonumber\\
&
\boldsymbol{\Gamma}_{F}^{uu}
-
\boldsymbol{\Gamma}_{F}^{uo}
(\boldsymbol{\Gamma}_{F}^{oo})^{-1}
\boldsymbol{\Gamma}_{F}^{ou}
\Big).
\label{s-eq:prediction-F}
\end{align}

\section{Laplace Approximation in the Collapsed Step}
\label{s-sec:LA}

The Laplace approximation for the collapsed step involves three main steps:
(1) obtaining the MAP estimate $\widehat{\mathbf{H}}$ via optimization,
(2) computing the Hessian matrix
$\nabla^{2}\{\operatorname{vec}(\widehat{\mathbf{H}})\}$
of the negative log-posterior at the MAP estimate, and
(3) sampling from the approximating Gaussian distribution
\[
\mathcal{N}\left(
\operatorname{vec}(\widehat{\mathbf{H}}),
\left[\nabla^{2}\{\operatorname{vec}(\widehat{\mathbf{H}})\}\right]^{-1}
\right).
\]
Here, we use $\nabla^{2}\{\operatorname{vec}(\mathbf{H})\}$ to denote the Hessian of the negative log-posterior with respect to $\operatorname{vec}(\mathbf{H})$.

The MAP estimate of $\mathbf{H}$ is obtained by solving
\begin{align*}
\widehat{\mathbf{H}}
&=
\arg\min_{\mathbf{H}}
\left\{-\log p(\mathbf{H}\mid\mathbf{Y})\right\} \\
&=
\arg\min_{\mathbf{H}}
\left\{
-\log g(\mathbf{Y}\mid\phi^{-1}(\mathbf{H}))
-\log p(\mathbf{H})
\right\}.
\end{align*}

We use the L-BFGS optimizer to solve this problem, which requires the gradient of the negative log-posterior:
\[
-\frac{\partial\log p(\mathbf{H}\mid\mathbf{Y})}{\partial\operatorname{vec}(\mathbf{H})}
=
-\frac{\partial\log g(\mathbf{Y}\mid\phi^{-1}(\mathbf{H}))}{\partial\operatorname{vec}(\mathbf{H})}
-
\frac{\partial\log p(\mathbf{H})}{\partial\operatorname{vec}(\mathbf{H})}.
\]

For the ALR parameterization used in MultiAddGPs, let
\(
\boldsymbol{\lambda}
=
\mathbf{Y}^{T}\mathbf{1}_{D},
\)
where $\lambda_n=\sum_{d=1}^{D}Y_{dn}$, and define
\[
\mathbf{J}
=
\operatorname{vec}\left[
\exp(\mathbf{H})
\oslash
\left\{
\mathbf{1}_{D-1}
\left(
\mathbf{1}_{N}
+
\exp(\mathbf{H})^{T}\mathbf{1}_{D-1}
\right)^{T}
\right\}
\right].
\]
Thus, $\mathbf{J}$ contains the multinomial probabilities corresponding to the first $D-1$ categories.

The likelihood contribution to the gradient is
\begin{align}
-\frac{\partial\log g(\mathbf{Y}\mid\phi^{-1}(\mathbf{H}))}{\partial\operatorname{vec}(\mathbf{H})}
&=
\operatorname{vec}
\left(
\mathbf{1}_{D-1}\boldsymbol{\lambda}^{T}
\right)
\odot\mathbf{J}
-
\operatorname{vec}
\left(
\mathbf{Y}_{1:(D-1),\cdot}
\right),
\label{s-eq:gradient_Y}
\end{align}
where $\odot$ denotes element-wise multiplication.

For the matrix-$t$ prior
$\mathbf{H}\sim T(\nu,\mathbf{M},\mathbf{V},\mathbf{A})$, define
\begin{align*}
\mathbf{S}
&=
\mathbf{I}_{D-1}
+
\mathbf{V}^{-1}
(\mathbf{H}-\mathbf{M})
\mathbf{A}^{-1}
(\mathbf{H}-\mathbf{M})^{T},\\
\mathbf{R}
&=
\mathbf{S}^{-1}\mathbf{V}^{-1},\\
\mathbf{C}
&=
\mathbf{A}^{-1}(\mathbf{H}-\mathbf{M})^{T},\\
c
&=
\frac{\nu+N+D-2}{2}=\frac{\zeta+N}{2}.
\end{align*}
The prior contribution to the gradient is then
\begin{align}
-\frac{\partial\log p(\mathbf{H})}{\partial\operatorname{vec}(\mathbf{H})}
&=
c\,
\operatorname{vec}
\left\{
(\mathbf{R}+\mathbf{R}^{T})\mathbf{C}^{T}
\right\}.
\label{s-eq:gradient_H}
\end{align}

Once the MAP estimate $\widehat{\mathbf{H}}$ has been obtained, we compute the Hessian of the negative log-posterior,
\[
\nabla^{2}\{\operatorname{vec}(\mathbf{H})\}
=
-\frac{\partial^{2}
\log g(\mathbf{Y}\mid\phi^{-1}(\mathbf{H}))}{\partial\operatorname{vec}(\mathbf{H})
 \partial\operatorname{vec}(\mathbf{H})^{T}}
-
\frac{\partial^{2}\log p(\mathbf{H})}{\partial\operatorname{vec}(\mathbf{H})
 \partial\operatorname{vec}(\mathbf{H})^{T}},
\]
and evaluate it at $\mathbf{H}=\widehat{\mathbf{H}}$.

Let $\mathbf{J}_{(n)}$ denote the $(D-1)$-dimensional subvector of $\mathbf{J}$ corresponding to sample $n$. The negative log-likelihood contribution to the Hessian is
\begin{align*}
-\frac{\partial^{2}
\log g(\mathbf{Y}\mid\phi^{-1}(\mathbf{H}))}{\partial\operatorname{vec}(\mathbf{H})
 \partial\operatorname{vec}(\mathbf{H})^{T}}
&=
\operatorname{diag}
\left(
\mathbf{W}^{(1)},\ldots,\mathbf{W}^{(N)}
\right),
\end{align*}
where
\[
\mathbf{W}^{(n)}
=
\lambda_n
\left\{
\operatorname{diag}(\mathbf{J}_{(n)})
-
\mathbf{J}_{(n)}\mathbf{J}_{(n)}^{T}
\right\}.
\]

For the matrix-$t$ contribution, define
\(
\mathbf{L}
=
\mathbf{C}\mathbf{R}\mathbf{C}^{T}
\otimes
\mathbf{R}^{T}.
\)
Then
\begin{align*}
-\frac{\partial^{2}\log p(\mathbf{H})}{\partial\operatorname{vec}(\mathbf{H})
 \partial\operatorname{vec}(\mathbf{H})^{T}}
={}&
c\Big[
\mathbf{A}^{-1}\otimes(\mathbf{R}+\mathbf{R}^{T})
-
(\mathbf{L}+\mathbf{L}^{T}) \\
&\quad -
\mathbf{T}_{N,D-1}
\left\{
(\mathbf{R}\mathbf{C}^{T}\otimes\mathbf{C}\mathbf{R}^{T})
+
(\mathbf{R}^{T}\mathbf{C}^{T}\otimes\mathbf{C}\mathbf{R})
\right\}
\Big],
\end{align*}
where $\mathbf{T}_{N,D-1}$ is the
$N(D-1)\times N(D-1)$ vec-transposition permutation matrix.
These gradient and Hessian expressions follow from the matrix-$t$ and ALR-parameterized multinomial derivatives in \citet{s-silverman2022bayesian}.

Finally, the Laplace approximation to the posterior
$p(\mathbf{H}\mid\mathbf{Y})$ is
\[
q(\mathbf{H}\mid\mathbf{Y})
=
\mathcal{N}\left(
\operatorname{vec}(\widehat{\mathbf{H}}),
\left[
\nabla^{2}\{\operatorname{vec}(\widehat{\mathbf{H}})\}
\right]^{-1}
\right).
\]
In practice, we use a Cholesky decomposition of the Hessian to sample from this Gaussian approximation without explicitly computing the inverse Hessian.

\section{Derivation of the posterior distribution of Matrix Normal}
\label{s-sec:matrix_normal}
\begin{theorem} 
If
\begin{align*}
\mathbf{Y} \mid \mathbf{\Lambda} &\sim \mathcal{MN}(\mathbf{\Lambda X}, \boldsymbol{\Sigma}, \mathbf{\Gamma}),\\
\mathbf{\Lambda} &\sim \mathcal{MN}(\mathbf{\Theta}, \boldsymbol{\Sigma}, \mathbf{Z}), 
\end{align*}
and \(\boldsymbol{\Sigma}\) is known, then the posterior of \(\mathbf{\Lambda}\) is given by:
\[
\mathbf{\Lambda} \mid \boldsymbol{\Sigma}, \mathbf{Y}
\sim \mathcal{MN}(\mathbf{\Theta}_N,\boldsymbol{\Sigma},\mathbf{Z}_N),
\]
where
\begin{align*}
\mathbf{Z}_N
&=(\mathbf{X}\mathbf{\Gamma}^{-1}\mathbf{X}^T+\mathbf{Z}^{-1})^{-1},\\
\mathbf{\Theta}_N
&=(\mathbf{Y}\mathbf{\Gamma}^{-1}\mathbf{X}^T+\mathbf{\Theta}\mathbf{Z}^{-1})\mathbf{Z}_N.
\end{align*}
\end{theorem}
\begin{proof}
Using the density function of the matrix normal distribution, we can write:
\[
\mathbf{\Lambda} \mid \mathbf{Y} \propto \exp\left[ -\frac{1}{2} \text{tr} \left\{ \boldsymbol{\Sigma}^{-1} (\mathbf{Y} - \mathbf{\Lambda} \mathbf{X}) \mathbf{\Gamma}^{-1} (\mathbf{Y} - \mathbf{\Lambda} \mathbf{X})^T \right\} \right] \times \exp\left[ -\frac{1}{2} \text{tr} \left\{ \boldsymbol{\Sigma}^{-1} (\mathbf{\Lambda} - \mathbf{\Theta}) \mathbf{Z}^{-1} (\mathbf{\Lambda} - \mathbf{\Theta})^T \right\} \right].
\]
Combining the exponents and expanding the term:
\begin{align*}
&\propto \exp\left[ -\frac{1}{2} \operatorname{tr}  \left\{\boldsymbol{\Sigma}^{-1} \left( \mathbf{Y} \mathbf{\Gamma}^{-1} \mathbf{Y}^T - \mathbf{\Lambda} \mathbf{X} \mathbf{\Gamma}^{-1} \mathbf{Y}^T - \mathbf{Y} \mathbf{\Gamma}^{-1} \mathbf{X}^T \mathbf{\Lambda}^T + \mathbf{\Lambda} \mathbf{X} \mathbf{\Gamma}^{-1} \mathbf{X}^T \mathbf{\Lambda}^T \right. \right. \right. \\
&\quad \left. \left. \left. + \mathbf{\Lambda} \mathbf{Z}^{-1} \mathbf{\Lambda}^T - \mathbf{\Theta} \mathbf{Z}^{-1} \mathbf{\Lambda}^T - \mathbf{\Lambda} \mathbf{Z}^{-1} \mathbf{\Theta}^T + \mathbf{\Theta} \mathbf{Z}^{-1} \mathbf{\Theta}^T \right) \right\} \right],
\end{align*}
\begin{align*}
& \propto \exp\left[ -\frac{1}{2} \text{tr} \left\{ \boldsymbol{\Sigma}^{-1} \left( -\mathbf{\Lambda} \mathbf{X} \mathbf{\Gamma}^{-1} \mathbf{Y}^T - \mathbf{Y} \mathbf{\Gamma}^{-1} \mathbf{X}^T \mathbf{\Lambda}^T + \mathbf{\Lambda} \mathbf{X} \mathbf{\Gamma}^{-1} \mathbf{X}^T \mathbf{\Lambda}^T + \mathbf{\Lambda} \mathbf{Z}^{-1} \mathbf{\Lambda}^T \right. \right. \right. \\
&\quad \left. \left. \left. -\mathbf{\Theta} \mathbf{Z}^{-1} \mathbf{\Lambda}^T - \mathbf{\Lambda} \mathbf{Z}^{-1} \mathbf{\Theta}^T \right) \right\} \right].
\end{align*}
Grouping like terms:
\begin{align*}
\mathbf{\Lambda} \mid \mathbf{Y}
&\propto \exp\Biggl( -\frac{1}{2} \text{tr}\Bigl[\boldsymbol{\Sigma}^{-1}
\Bigl\{\mathbf{\Lambda}(\mathbf{X}\mathbf{\Gamma}^{-1}\mathbf{X}^T+\mathbf{Z}^{-1})\mathbf{\Lambda}^T\\
&\qquad -\mathbf{\Lambda}(\mathbf{X}\mathbf{\Gamma}^{-1}\mathbf{Y}^T+\mathbf{Z}^{-1}\mathbf{\Theta}^T)\\
&\qquad -(\mathbf{Y}\mathbf{\Gamma}^{-1}\mathbf{X}^T+\mathbf{\Theta}\mathbf{Z}^{-1})\mathbf{\Lambda}^T
\Bigr\}\Bigr]\Biggr),
\end{align*}
\begin{align*}
&\propto \exp\left( -\frac{1}{2} \operatorname{tr} \left[ \boldsymbol{\Sigma}^{-1} \left\{ \mathbf{\Lambda} - (\mathbf{Y} \mathbf{\Gamma}^{-1} \mathbf{X}^{T} + \mathbf{\Theta} \mathbf{Z}^{-1}) (\mathbf{X} \mathbf{\Gamma}^{-1} \mathbf{X}^{T} + \mathbf{Z}^{-1})^{-1} \right\} \right. \right. \\
&\quad \left. \left. \times (\mathbf{X} \mathbf{\Gamma}^{-1} \mathbf{X}^{T} + \mathbf{Z}^{-1}) \left\{ \mathbf{\Lambda} - (\mathbf{Y} \mathbf{\Gamma}^{-1} \mathbf{X}^{T} + \mathbf{\Theta} \mathbf{Z}^{-1}) (\mathbf{X} \mathbf{\Gamma}^{-1} \mathbf{X}^{T} + \mathbf{Z}^{-1})^{-1} \right\}^{T} \right] \right),
\end{align*}
which implies that
\[
\mathbf{\Lambda} \mid \boldsymbol{\Sigma}, \mathbf{Y}
\sim \mathcal{MN}(\mathbf{\Theta}_N,\boldsymbol{\Sigma},\mathbf{Z}_N).
\]
\end{proof}
Note that in the special case where \(\mathbf{X} = \mathbf{I}\), i.e., a model of the form:
\[
\mathbf{Y} \mid \mathbf{\Lambda} \sim \mathcal{MN}(\mathbf{\Lambda}, \boldsymbol{\Sigma}, \mathbf{\Gamma}),
\]
\[
\mathbf{\Lambda} \sim \mathcal{MN}(\mathbf{\Theta}, \boldsymbol{\Sigma}, \mathbf{Z}),
\]
then the above result simplifies to
\[
\mathbf{\Lambda} \mid \boldsymbol{\Sigma}, \mathbf{Y} \sim \mathcal{MN}\left\{ (\mathbf{Y} \mathbf{\Gamma}^{-1} + \mathbf{\Theta} \mathbf{Z}^{-1}) (\mathbf{\Gamma}^{-1} + \mathbf{Z}^{-1})^{-1}, \boldsymbol{\Sigma}, (\mathbf{\Gamma}^{-1} + \mathbf{Z}^{-1})^{-1} \right\}.
\]

\section{Back-sampling}
\label{s-sec:backsampling}

For each posterior sample, the back-sampling algorithm proceeds by sampling
\(p(\mathbf{B} \mid \mathbf{F}, \boldsymbol{\Sigma})\) and then sequentially
sampling
\(p(\mathbf{F}^{(1)} \mid \mathbf{B}, \mathbf{F}, \boldsymbol{\Sigma}), \dots,
p(\mathbf{F}^{(K-1)} \mid \mathbf{F}^{(1)}, \dots,
\mathbf{F}^{(K-2)}, \mathbf{B}, \mathbf{F}, \boldsymbol{\Sigma})\).
Note that the value of the final component \(\mathbf{F}^{(K)}\) is uniquely
determined by \(\mathbf{F}, \mathbf{B}, \mathbf{F}^{(1)}, \dots,
\mathbf{F}^{(K-1)}\). Thus, a sample of \(\mathbf{F}^{(K)}\) from the joint
distribution is obtained by
\[
\mathbf{F}^{(K)}
=
\mathbf{F}
-
\mathbf{B}\mathbf{X}
-
\sum_{k=1}^{K-1}\mathbf{F}^{(k)}.
\]
The pseudocode for the back-sampling algorithm is provided below.

First, we sample from the posterior
\(p(\mathbf{B} \mid \mathbf{F}, \boldsymbol{\Sigma})\).
To do this, we marginalize over
\(\mathbf{F}^{(1)}, \dots, \mathbf{F}^{(K)}\) and reparameterize as
\[
\mathbf{B}^{\epsilon}
:=
\mathbf{F}
-
\sum_{k=1}^{K}
\boldsymbol{\Theta}^{(k)}(\mathbf{Z}^{(k)}).
\]

This leads to the following conjugate Bayesian model for \(\mathbf{B}\):

\begin{align*}
\mathbf{B}^{\epsilon}
&\sim
\mathcal{MN}
\left(
\mathbf{B}\mathbf{X},
\boldsymbol{\Sigma},
\boldsymbol{\Gamma}^{*}
\right),\\
\mathbf{B}
&\sim
\mathcal{MN}
\left(
\boldsymbol{\Theta}^{(0)},
\boldsymbol{\Sigma},
\boldsymbol{\Gamma}^{(0)}
\right),
\end{align*}

where
\(
\boldsymbol{\Gamma}^{*}
=
\sum_{k=1}^{K}
\boldsymbol{\Gamma}^{(k)}(\mathbf{Z}^{(k)}),
\)
and \(\boldsymbol{\Theta}^{(0)}\) and
\(\boldsymbol{\Gamma}^{(0)}\) denote the prior mean and covariance of
\(\mathbf{B}\), respectively. Using Section~\ref{s-sec:matrix_normal}, the
posterior is
\[
\begin{aligned}
\mathbf{B}
\mid
\mathbf{B}^{\epsilon},\boldsymbol{\Sigma}
\sim
\mathcal{MN}
&
\Bigl(
\Bigl[
\mathbf{B}^{\epsilon}(\boldsymbol{\Gamma}^{*})^{-1}\mathbf{X}^{T}
+
\boldsymbol{\Theta}^{(0)}
(\boldsymbol{\Gamma}^{(0)})^{-1}
\Bigr] \\
& 
\times \Bigl[
\mathbf{X}(\boldsymbol{\Gamma}^{*})^{-1}\mathbf{X}^{T}
+
(\boldsymbol{\Gamma}^{(0)})^{-1}
\Bigr]^{-1},
\boldsymbol{\Sigma}, \\
&
\Bigl[
\mathbf{X}(\boldsymbol{\Gamma}^{*})^{-1}\mathbf{X}^{T}
+
(\boldsymbol{\Gamma}^{(0)})^{-1}
\Bigr]^{-1}
\Bigr).
\end{aligned}
\]

We use a similar approach to sample each nonlinear component
\(p(\mathbf{F}^{(1)} \mid \mathbf{B}, \mathbf{F}, \boldsymbol{\Sigma}),
\dots,
p(\mathbf{F}^{(K-1)} \mid
\mathbf{F}^{(1)}, \dots, \mathbf{F}^{(K-2)},
\mathbf{B}, \mathbf{F}, \boldsymbol{\Sigma})\).
For each \(k\in\{1,\dots,K-1\}\), we reparameterize the model as
\[
\mathbf{F}^{\epsilon_k}
=
\mathbf{F}
-
\mathbf{B}\mathbf{X}
-
\sum_{i=1}^{k-1}\mathbf{F}^{(i)}
-
\sum_{j=k+1}^{K}
\boldsymbol{\Theta}^{(j)}(\mathbf{Z}^{(j)}).
\]

Similar to \(\mathbf{B}^{\epsilon}\),
\(\mathbf{F}^{\epsilon_k}\) leads to the following conjugate Bayesian
model for the finite-dimensional value of \(\mathbf{f}^{(k)}\) evaluated
at \(\mathbf{Z}^{(k)}\):
\begin{align*}
\mathbf{F}^{\epsilon_k}
&\sim
\mathcal{MN}
\left(
\mathbf{F}^{(k)},
\boldsymbol{\Sigma},
\boldsymbol{\Gamma}^{*}_{k}
\right),\\
\mathbf{F}^{(k)}
&\sim
\mathcal{MN}
\left(
\boldsymbol{\Theta}^{(k)}(\mathbf{Z}^{(k)}),
\boldsymbol{\Sigma},
\boldsymbol{\Gamma}^{(k)}(\mathbf{Z}^{(k)})
\right),
\end{align*}
where
\(
\boldsymbol{\Gamma}^{*}_{k}
=
\sum_{j=k+1}^{K}
\boldsymbol{\Gamma}^{(j)}(\mathbf{Z}^{(j)}).
\)

Again, using Section~\ref{s-sec:matrix_normal},
\[
\begin{aligned}
\mathbf{F}^{(k)}
\mid
\boldsymbol{\Sigma},\mathbf{F}^{\epsilon_k}
\sim
\mathcal{MN}
\Bigg(
&
\left[
\mathbf{F}^{\epsilon_k}
(\boldsymbol{\Gamma}^{*}_{k})^{-1}
+
\boldsymbol{\Theta}^{(k)}(\mathbf{Z}^{(k)}) \{\boldsymbol{\Gamma}^{(k)}(\mathbf{Z}^{(k)})\}^{-1}
\right]\\
&\times
\left[
(\boldsymbol{\Gamma}^{*}_{k})^{-1}
+
\left\{
\boldsymbol{\Gamma}^{(k)}(\mathbf{Z}^{(k)})
\right\}^{-1}
\right]^{-1},
\boldsymbol{\Sigma},\\
&
\left[
(\boldsymbol{\Gamma}^{*}_{k})^{-1}
+
\left\{
\boldsymbol{\Gamma}^{(k)}(\mathbf{Z}^{(k)})
\right\}^{-1}
\right]^{-1}
\Bigg).
\end{aligned}
\]

Finally, the last component is uniquely determined as
\[
\mathbf{F}^{(K)}
=
\mathbf{F}
-
\mathbf{B}\mathbf{X}
-
\sum_{k=1}^{K-1}\mathbf{F}^{(k)}.
\]

\begin{algorithm}
\caption{Back-sampling}
\label{s-algorithm:BS}

\parbox[t]{\linewidth}{%
\raggedright
\textbf{Input:}
\(\mathbf{X},\mathbf{Z}^{(1)},\dots,\mathbf{Z}^{(K)}\) are the model
covariates;
\(\{\mathbf{F},\boldsymbol{\Sigma}\}\) is one sample from the uncollapsed
step;
\(\Delta=
\{\boldsymbol{\Theta}^{(0)},\dots,\boldsymbol{\Theta}^{(K)},
\boldsymbol{\Gamma}^{(0)},\dots,\boldsymbol{\Gamma}^{(K)}\}\)
is the set of prior inputs.\\
\textbf{Output:}
one sample
\(\{\mathbf{B},\mathbf{F}^{(1)},\dots,\mathbf{F}^{(K)}\}\).
}

\begin{algorithmic}[1]

\State
\(
\mathbf{B}^{\epsilon}
=
\mathbf{F}
-
\sum_{k=1}^{K}
\boldsymbol{\Theta}^{(k)}(\mathbf{Z}^{(k)})
\)

\State
\(
\boldsymbol{\Gamma}^{*}
=
\sum_{k=1}^{K}
\boldsymbol{\Gamma}^{(k)}(\mathbf{Z}^{(k)})
\)

\State
\(
\boldsymbol{\Gamma}_{N}^{(0)}
=
\left[
\mathbf{X}(\boldsymbol{\Gamma}^{*})^{-1}\mathbf{X}^{T}
+
(\boldsymbol{\Gamma}^{(0)})^{-1}
\right]^{-1}
\)

\State
\(
\boldsymbol{\Theta}_{N}^{(0)}
=
\left[
\mathbf{B}^{\epsilon}(\boldsymbol{\Gamma}^{*})^{-1}\mathbf{X}^{T}
+
\boldsymbol{\Theta}^{(0)}(\boldsymbol{\Gamma}^{(0)})^{-1}
\right]
\boldsymbol{\Gamma}_{N}^{(0)}
\)

\State
\(
\mathbf{B}
\mid
\mathbf{B}^{\epsilon},\boldsymbol{\Sigma}
\sim
\mathcal{MN}
\left(
\boldsymbol{\Theta}_{N}^{(0)},
\boldsymbol{\Sigma},
\boldsymbol{\Gamma}_{N}^{(0)}
\right)
\)

\For{$k=1$ to $K-1$}

\State
\(
\mathbf{F}^{\epsilon_k}
=
\mathbf{F}
-
\mathbf{B}\mathbf{X}
-
\sum_{i=1}^{k-1}\mathbf{F}^{(i)}
-
\sum_{j=k+1}^{K}
\boldsymbol{\Theta}^{(j)}(\mathbf{Z}^{(j)})
\)

\State
\(
\boldsymbol{\Gamma}^{*}_{k}
=
\sum_{j=k+1}^{K}
\boldsymbol{\Gamma}^{(j)}(\mathbf{Z}^{(j)})
\)

\State
\(
\boldsymbol{\Gamma}_{N}^{(k)}
=
\left[
(\boldsymbol{\Gamma}^{*}_{k})^{-1}
+
\left\{
\boldsymbol{\Gamma}^{(k)}(\mathbf{Z}^{(k)})
\right\}^{-1}
\right]^{-1}
\)

\State
\(
\boldsymbol{\Theta}_{N}^{(k)}
=
\left[
\mathbf{F}^{\epsilon_k}
(\boldsymbol{\Gamma}^{*}_{k})^{-1}
+
\boldsymbol{\Theta}^{(k)}(\mathbf{Z}^{(k)}) \{\boldsymbol{\Gamma}^{(k)}(\mathbf{Z}^{(k)})\}^{-1}
\right]
\boldsymbol{\Gamma}_{N}^{(k)}
\)

\State
\(
\mathbf{F}^{(k)}
\mid
\boldsymbol{\Sigma},\mathbf{F}^{\epsilon_k}
\sim
\mathcal{MN}
\left(
\boldsymbol{\Theta}_{N}^{(k)},
\boldsymbol{\Sigma},
\boldsymbol{\Gamma}_{N}^{(k)}
\right)
\)

\EndFor

\State
\(
\mathbf{F}^{(K)}
=
\mathbf{F}
-
\mathbf{B}\mathbf{X}
-
\sum_{k=1}^{K-1}\mathbf{F}^{(k)}
\)

\end{algorithmic}
\end{algorithm}

\section{Simulation study}
\label{s-sec:simulation}

\subsection{Comparison to Lgpr and GAM}
\label{s-subsec:sim-lgpr-gam}
The simulation in main text Section \ref{m-sec:simulation} follows the model:
\begin{align*}
\mathbf{F}^{(\text{Intervention})} &\sim \mathcal{MN}(0,\boldsymbol{\Sigma}, \mathbf{\Gamma}^{(\text{Intervention})}),\\
\mathbf{F}^{(\text{General})} &\sim \mathcal{MN}(0,\boldsymbol{\Sigma}, \mathbf{\Gamma}^{(\text{General})}), \\
\mathbf{F}_{\cdot n} &= 0.4 + x_{n}^{(\text{Batch})} +  \mathbf{f}^{(\text{General})}(t_n) + \mathbf{f}^{(\text{Intervention})}(w_n),\\
\mathbf{H}_{\cdot n} &\sim  \mathcal{N}(\mathbf{F}_{\cdot n}, \boldsymbol{\Sigma}),\\
\boldsymbol{\Pi}_{\cdot n} &= ALR^{-1}(\mathbf{H}_{\cdot n}),\\
\mathbf{Y}_{\cdot n} &\sim \text{Multinomial}(\lambda_n, \boldsymbol{\Pi}_{\cdot n}).
\end{align*}
Here, \(w_n=w(t_n)\) is a transformed skewed Gaussian of input \(t_n\), used in the non-stationary kernel model:
\begin{align*}
z_n &= \frac{t_n - m}{s},\\
w_n &= \exp\left(-\frac{1}{2}z_n^2\right) \times \frac{1}{1+ \exp(-\alpha z_n)},
\end{align*}
where $m$ determines the peak location of the distribution, $s$ controls its width, and $\alpha$ is the skew parameter. When 
$\alpha = 0$, the function reduces to a standard Gaussian; positive $\alpha$ induces a heavier right tail, while negative $\alpha$ results in a heavier left tail. The resulting values \(w_n\) were then normalized to \([-1,1]\) before use in the intervention kernel. In this simulation, the covariance matrix $\boldsymbol{\Sigma}$ is set with off-diagonal elements of 0.5 and diagonal elements of 1. The non-stationary kernel is defined as
\[
\Gamma^{(\text{Intervention})}(w_n,w_{n'})
=
4 \times \text{exp}\left\{
-\frac{(w_n-w_{n'})^2}{2 \times 3^2}
\right\},
\]
while the general kernel is modeled on standardized time as
\[
\Gamma^{(\text{General})}(t_n,t_{n'})
=
\text{exp}\left\{
-\frac{(t_n-t_{n'})^2}{2 \times 0.5^2}
\right\}.
\]
In all simulations, we centered the batch, general, and intervention component estimates 
after fitting to impose a sum-to-zero constraint. For each \((D,N)\) configuration,
100 replicate datasets were generated. For each method, each replicate was
allocated 4 CPUs and 96 GB RAM on a single compute node, with a 48-hour wall-clock time 
limit. In practice, this limit primarily affected the Lgpr models, which were therefore 
evaluated only on a subset of the \((D,N)\) settings. Specifically, Gaussian Lgpr was
evaluated for \(D=3\) with \(N\in\{20,50, 70, 100\}\) and for \(D=10\) with \(N=50\), 
while Negative Binomial Lgpr was evaluated for \(D=3\) with \(N\in\{20,50\}\). 
In contrast, MultiAddGPs and GAM completed all tested \((D,N)\) configurations within 
the allotted time.

We recorded elapsed wall-clock time for the final model-fitting call on each replicate.
 For MultiAddGPs, this is the final \texttt{fido::basset()} fit.
  For GAM and both Lgpr variants, we fitted one taxon and multiplied its recorded fitting time by \(D\) to
   estimate the time required to fit all taxa sequentially. Figure~2 shows the mean of these per-replicate
    fitting-time estimates across 100 seeds for each setting.

For MultiAddGPs, we fit a multinomial logistic-normal model using \texttt{fido::basset}, 
decomposing the latent ALR mean into a linear batch effect and two mean-zero Gaussian process 
components representing the general and intervention effects. The first two covariates were 
included as linear effects (\texttt{linear = 1:2}) with prior mean $\boldsymbol{\Theta}_{\text{linear}}=\mathbf{0}_{(D-1)\times 2}$ 
and prior covariance $\boldsymbol{\Gamma}_{\text{linear}}$ with diagonal entries $1+10^{-6}$ and 
off-diagonal entries $10^{-6}$. The nonlinear components were assigned squared exponential kernels 
for the general and intervention effects. We set $\zeta=D+100$, used \texttt{n\_samples = 2000}, 
\texttt{max\_iter = 1000}, and \texttt{jitter = 1e-4}, and initialized start point at $\mathrm{ALR}(\mathbf{Y}+0.65)$. 
Kernel hyperparameters $(\sigma_{\text{intervention}}, \rho_{\text{intervention}}, \sigma_{\text{general}}, 
\rho_{\text{general}})$ were selected by Bayesian optimization of a penalized log marginal likelihood with weak 
Gaussian regularization.

Since Lgpr utilizes predefined kernels, we selected kernel configurations that most closely matched the simulated model structure. Lgpr-NB and Lgpr-Gaussian used the same covariate and kernel specification and differed only in their observation models: Lgpr-NB was fit to the count outcomes using a Negative Binomial likelihood, whereas Lgpr-Gaussian was fit to the response $\mathrm{CLR}(\mathbf{Y}+0.65)$ using a Gaussian likelihood. Unless otherwise noted, we use ``Lgpr'' below to refer to both variants. Models were fitted using the \texttt{lgp} function with a structured formula. A temporal Gaussian process, $gp(t_n)$, was included to capture the background temporal trend, the sum-to-zero kernel $zs(x_n^{\text{batch}})$ was used for the categorical batch effect, and $gp\_ns(t_n)$ was used to model the non-stationary intervention effect. Lgpr-Gaussian was run for 6,500 iterations, whereas Lgpr-NB was run for 6,500 iterations at \(N=20\) and 8,000 iterations at \(N=50\). All models used 4 MCMC chains, with a refresh interval of 3,250 or 4,000 in the current implementation. To improve sampling efficiency, the control settings used an adaptation delta of 0.999 or 0.9999 and a maximum tree depth of 11. Most saved Lgpr-NB fits used \texttt{adapt\_delta = 0.9999}.

For comparisons on the CLR scale, we converted Lgpr-NB posterior predictions for taxon 2 using the observed counts of the remaining taxa, with a pseudocount of 0.65 applied to those counts. These observed counts were held fixed across posterior draws, so the resulting intervals are conditional on the observed counts of the other taxa. Estimation accuracy and interval coverage were evaluated for the same taxon-2 CLR coordinate across all methods.

Lgpr models were run for prespecified numbers of iterations, and effective sample size (ESS) was assessed after fitting. For each fit, we calculated the median ESS across parameters. Median ESS was at least 2,000 in 499 of 500 Lgpr-Gaussian fits and 197 of 200 Lgpr-NB fits.

For the GAM baseline, we fit an additive model to the response $\mathrm{CLR}(\mathbf{Y}+0.65)$ with three components: a parametric batch term and two smooth functions. Batch was included as a categorical fixed effect to capture $x_n^{(\text{batch})}$. To represent the broad temporal trend over standardized time $t_n$, corresponding to $\mathbf{f}^{(\text{General})}(t_n)$, we used a smooth spline with a small basis dimension ($k=5$), enforcing a deliberately smooth, low-complexity trajectory. To model the intervention effect, we fit a second smooth over standardized time using an adaptive smooth basis ($k=12$), allowing the effective smoothness to vary across the domain rather than imposing a single global length scale. Smoothing penalties were estimated automatically within \texttt{mgcv}, so the GAM comparison did not require an explicit kernel specification.

We further assessed the sensitivity of MultiAddGPs to the Gaussian process
hyperparameters using the simulation setting with \(D=3\) taxa and \(N=50\)
samples. Starting from the baseline hyperparameter specification, we perturbed
the magnitude \(\sigma\) and length-scale \(\rho\) parameters of the general
and intervention components one at a time by \(\pm10\%\). For each perturbation,
we evaluated both the root mean squared error (RMSE) and empirical coverage of
the estimated batch, general temporal, intervention, and cumulative effects
across 10 replicate datasets.

\begin{figure}
    \centering
    \includegraphics[width=1\linewidth]{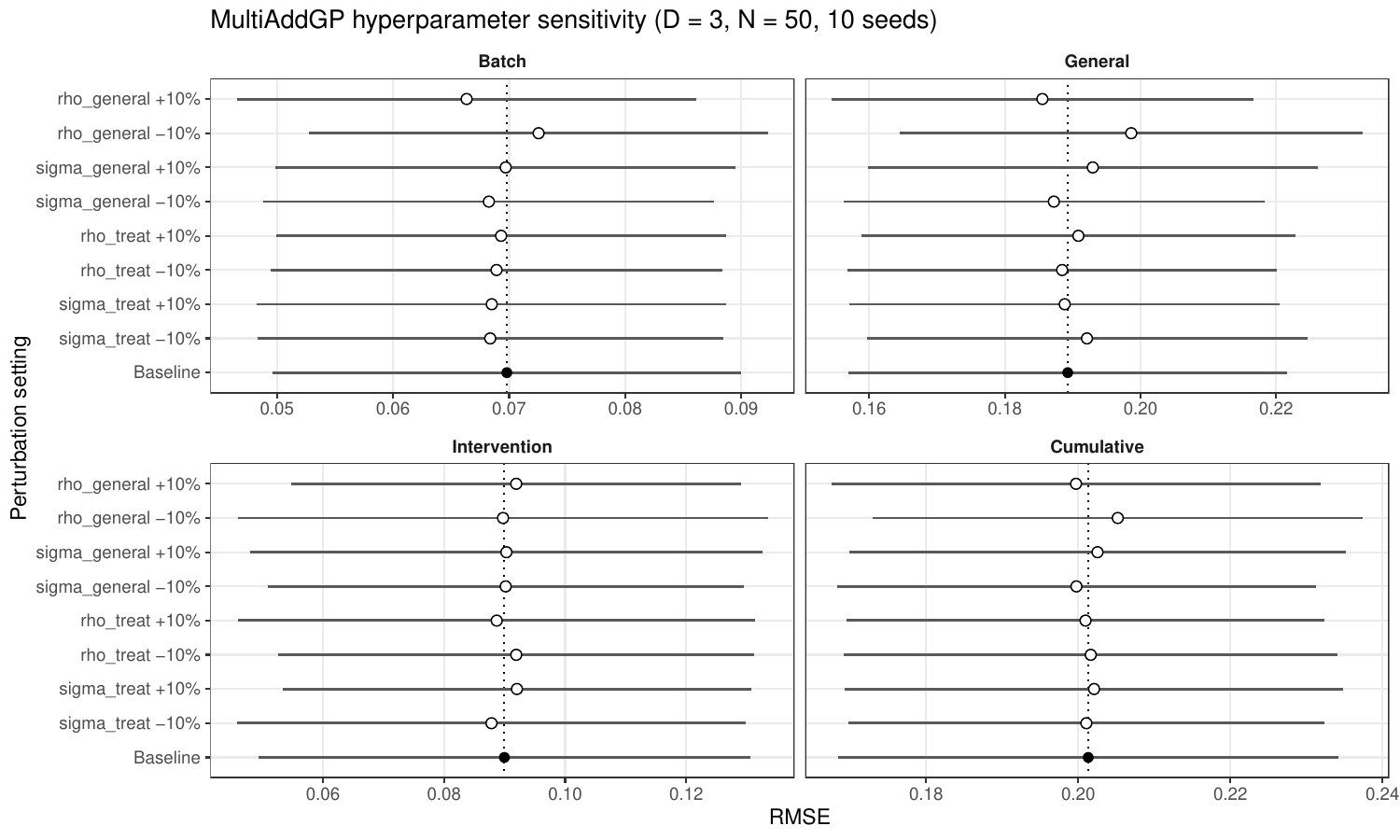}
    \caption{Sensitivity of MultiAddGP inference to GP hyperparameter
specification.}
    \label{s-fig:sim-hyper-sensitivity-rmse}
\end{figure}

\begin{figure}
    \centering
    \includegraphics[width=1\linewidth]{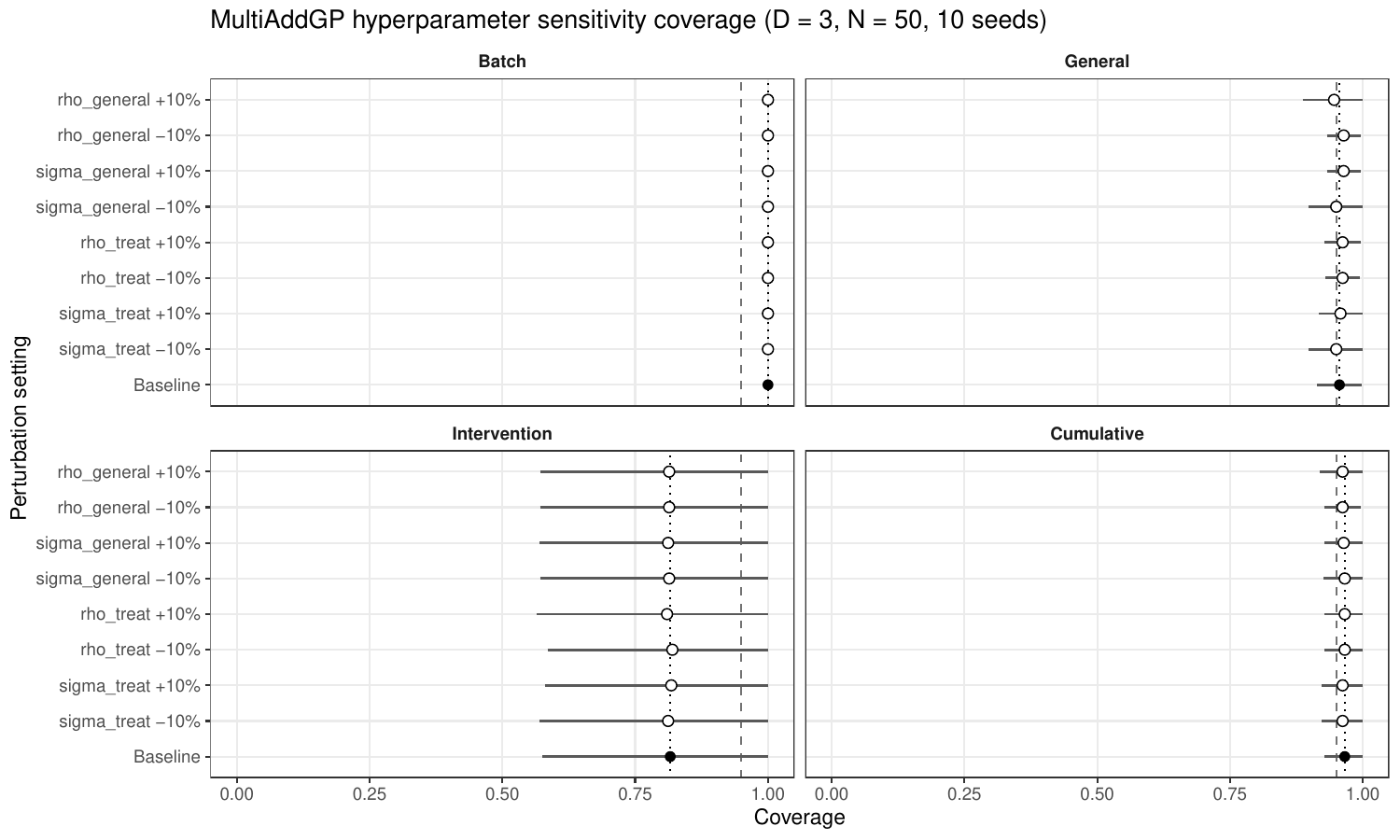}
    \caption{Sensitivity of MultiAddGP inference to GP hyperparameter
specification.}
    \label{s-fig:sim-hyper-sensitivity-coverage}
\end{figure}

As shown in Figure~\ref{s-fig:sim-hyper-sensitivity-rmse} and Figure~\ref{s-fig:sim-hyper-sensitivity-coverage}, both RMSE and empirical
coverage remained similar to the baseline specification across the considered
hyperparameter perturbations. These results suggest that both estimation
accuracy and uncertainty quantification are robust to moderate changes in the
GP hyperparameters in this simulation setting.

\subsection{Comparison to NAddGPs}
\label{s-subsec:sim-Naddgps}

We simulated a longitudinal microbiome study with \(D=4\) taxa observed over \(N=600\) time points. We simulated
microbial composition influenced by batch effects, daily periodicity (e.g., circadian rhythm;~\cite{s-heddes2022intestinal}),
and longer-term trends. We created a new method \textit{Normal Additive GPs
(\textit{NAddGPs})} which assumes the data are transformed Gaussian. NAddGPs are nearly identical to MultiAddGPs but assume each \(\mathbf{H}_{\cdot n}\) is observed and equal to
\(\mathbf{H}_{\cdot n}=\phi(\mathbf{Y}_{\cdot n} + 0.65 \cdot \mathbf{1}_{D})\) and therefore proceed directly to
the uncollapse step, skipping the collapsed step of our CU sampler.

To evaluate the implementation and investigate the behavior of the MultiAddGPs model, let \(t_{n}\) denote the time at which sample \(n\) was obtained; we simulated a synthetic microbial community time-series based on the following model:
\begin{align*}
    \mathbf{F}^{(\text{periodic})} &\sim \mathcal{MN}(0,\boldsymbol{\Sigma}, \mathbf{\Gamma}^{(\text{periodic})}),\\
    \mathbf{F}^{(\text{trend})} &\sim \mathcal{MN}(0,\boldsymbol{\Sigma}, \mathbf{\Gamma}^{(\text{trend})}),\\
    \mathbf{F}_{\cdot n} &= \mathbf{b}_{0} +  \mathbf{b}_{1}x_{n}^{(\text{batch})}
    + \mathbf{f}^{(\text{periodic})}(t_n) + \mathbf{f}^{(\text{trend})}(t_n),\\
    \boldsymbol{\Pi}_{\cdot n} &= \mathrm{ALR}^{-1}(\mathbf{H}_{\cdot n}),\\
    \mathbf{Y}_{\cdot n} &\sim \text{Multinomial}(\lambda_n, \boldsymbol{\Pi}_{\cdot n}).
\end{align*}
Here, we set \(\boldsymbol{\Sigma}\) as a covariance matrix with off-diagonal elements of \(0.9\) and diagonal
elements of \(1.5\). The periodic kernel is defined as
\(\Gamma^{(\text{periodic})}(t,t') = 16 \exp\!\left\{-2 \sin^2\!\left(\pi |t - t'|/30\right)/25^2\right\}\),
while the trend kernel is modeled as
\(\Gamma^{(\text{trend})}(t,t') = \exp\!\left\{-(t - t')^2/(2 \times 30^2)\right\}\).
After obtaining posterior samples from the MultiAddGPs model, we applied a sum-to-zero constraint to facilitate
model identification.

For inference, we specified MultiAddGPs and NAddGPs as
\(\mathbf{F}_{\cdot n}=\mathbf{b}_{0} + \mathbf{b}_{1}x^{(\text{batch})}_{n}
+ \mathbf{f}^{(\text{periodic})}(t_{n})+\mathbf{f}^{(\text{trend})}(t_{n})\). For covariates, we set
\(\mathbf{X}_{\cdot n}=[1\; x^{(\text{batch})}_{n}]^{T}\),
\(\mathbf{Z}^{(\text{periodic})}_{\cdot n}=t_{n}\), and \(\mathbf{Z}^{(\text{trend})}_{\cdot n}=t_{n}\). For
priors we set \(\mathbf{B} = [\mathbf{b}_{0} = 2.7 ; \mathbf{b_{1}}= 1] \). Both
\(\mathbf{f}^{(\text{periodic})}\) and \(\mathbf{f}^{(\text{trend})}\) were given matrix-normal process priors
with mean function \(\boldsymbol{\Theta}^{(k)}=\mathbf{0}\). A periodic kernel
\(K_{\text{period}}(t, t') = \sigma_{\text{period}}^2 \exp\left\{-2
\sin^2\left(\pi |t - t'|/p\right)/\rho_{\text{period}}^2\right\}\) was used in the prior for \(\mathbf{f}^{(\text{periodic})}\)
and a squared exponential
\(K_{\text{trend}}(t,t') = \sigma_{\text{trend}}^2 \exp\{-(t-t')^2/(2 \rho_{\text{trend}}^2)\}\) was used
for \(\mathbf{f}^{(\text{trend})}\). Hyperparameters
\(\Omega=\{\sigma_{\text{period}},\rho_{\text{period}}, \sigma_{\text{trend}},\rho_{\text{trend}}\}\) were
selected using penalized marginal likelihood with $p$ fixed at 30.

\begin{figure}
    \centering
    \includegraphics[width=1\linewidth]{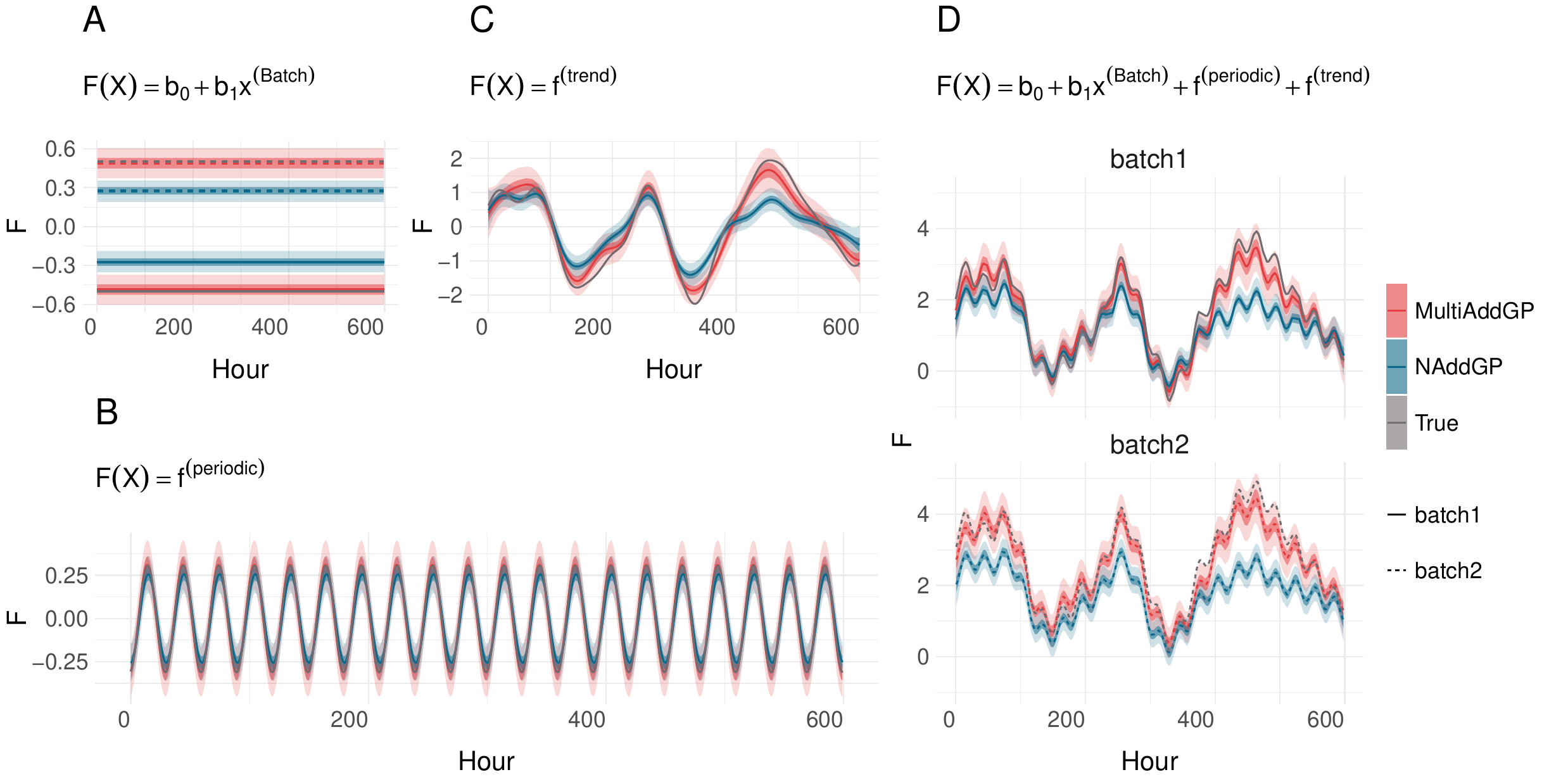}
    \caption{MultiAddGPs successfully decompose simulated microbiome time-series on Taxa 1.}
    \label{s-fig: Naddgps_taxa1}
\end{figure}

\begin{figure}
    \centering
    \includegraphics[width=1\linewidth]{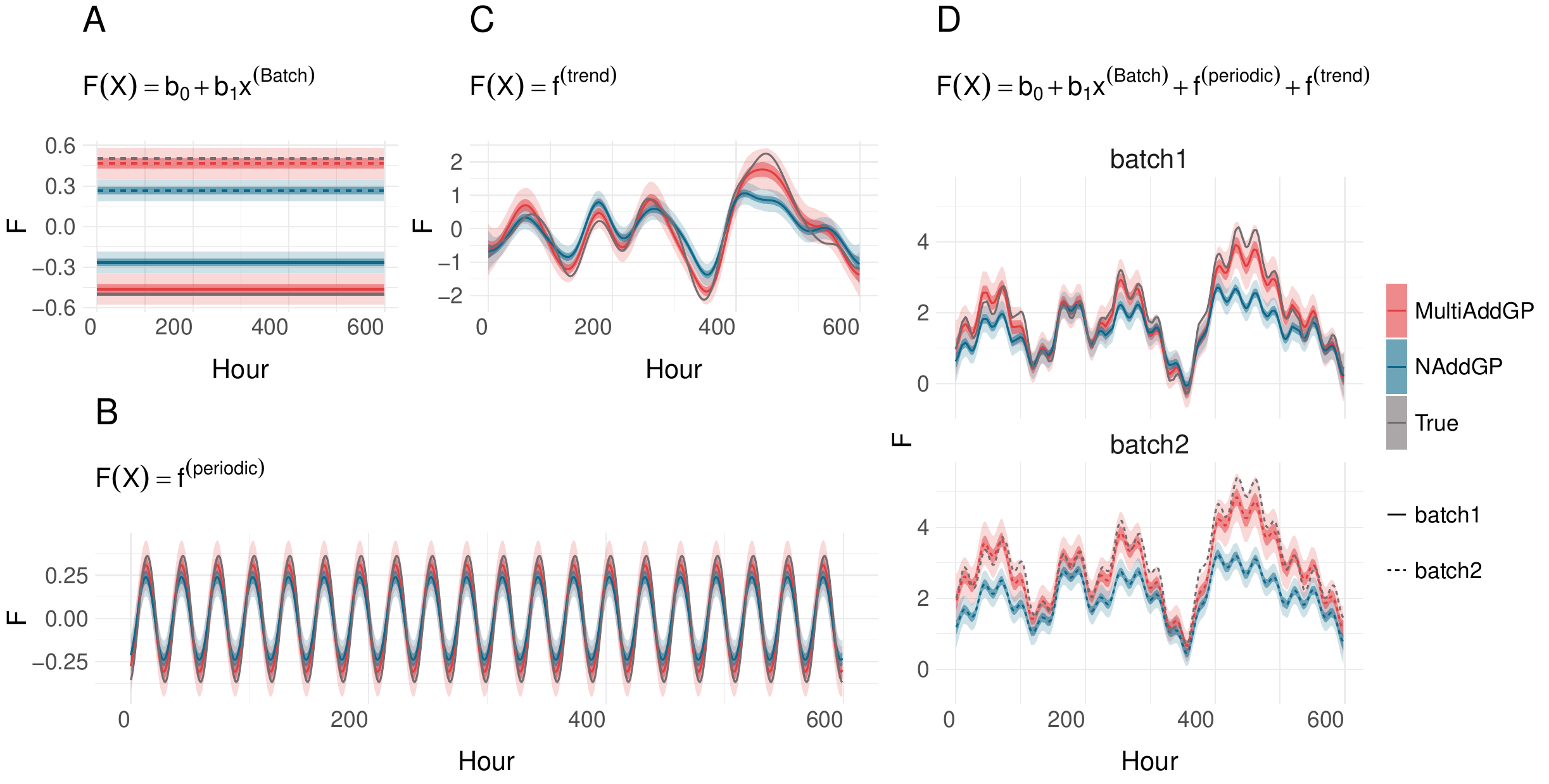}
    \caption{MultiAddGPs successfully decompose simulated microbiome time-series on Taxa 2.}
    \label{s-fig: Naddgps_taxa2}
\end{figure}

\begin{figure}
    \centering
    \includegraphics[width=1\linewidth]{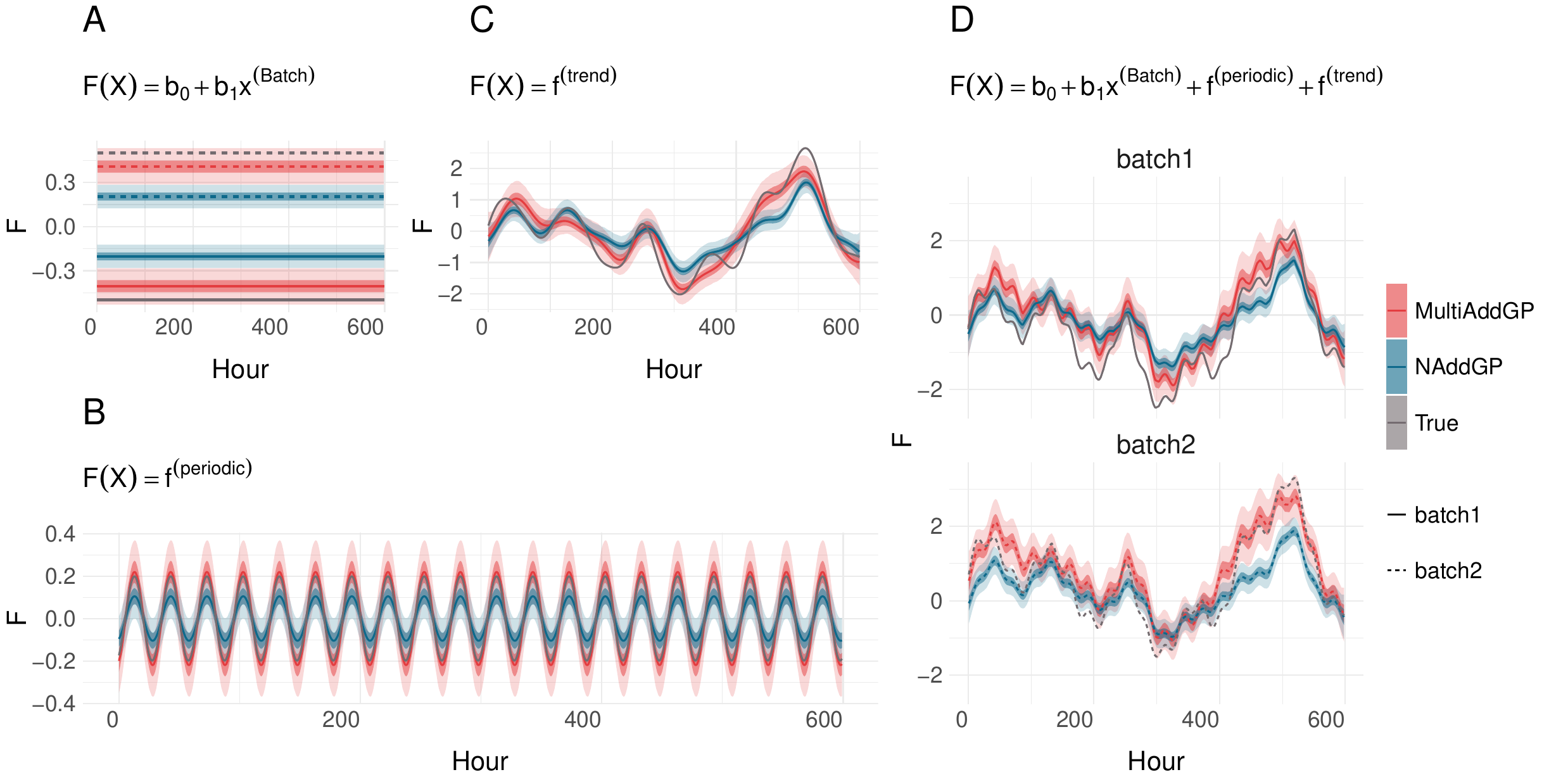}
    \caption{MultiAddGPs successfully decompose simulated microbiome time-series on Taxa 3.}
    \label{s-fig: Naddgps_taxa3}
\end{figure}

We compared posterior estimates from the MultiAddGPs and NAddGPs models to emphasize the importance of modeling
uncertainty due to counting. Figure~\ref{s-fig: Naddgps_taxa1}, \ref{s-fig: Naddgps_taxa2}, and \ref{s-fig: Naddgps_taxa3} show that the MultiAddGPs almost perfectly
recovered the true decomposition, whereas the NAddGPs substantially underestimated the amplitude of the periodic
component and long-term trend. 

All implementations were compiled and run using gcc version 9.1.0 and R version 4.3.2. All replicates of the
simulated count data were supplied to the various implementations independently and the models were fit on
identical hardware, allotted 64GB RAM, 4 cores, and restricted to a 48-hour upper limit on run-time.

\section{Details for the Two Longitudinal Microbiome Applications}
\label{s-sec:applications}

\subsection{Artificial-gut analysis}
\label{s-subsec:artificial_data}

We applied MultiAddGPs to a previously published longitudinal study of four artificial human gut models~\citep{s-silverman2018dynamic}. The study contains over 500 microbiome samples collected at irregular time points from \(4\) physically isolated vessels. A key experimental feature is a feed-starvation intervention applied to Vessels 1 and 2 over Days 11--12, while Vessels 3 and 4 received no intervention and therefore act as controls. Previous analyses~\citep{s-silverman2018dynamic,s-silverman2022bayesian} documented prolonged post-starvation recovery, but could not cleanly separate long-term temporal drift from intervention-specific dynamics. Our goal is to disentangle these overlapping processes and clarify the recovery phase.

We followed the preprocessing procedure of \citet{s-silverman2018dynamic},
yielding $D=10$ bacterial families and $N=537$ total samples across the four vessels. We analyzed CLR/ALR coordinates under a multinomial-logistic normal model as in the MultiAddGPs framework.

\paragraph{Model specification}

We represent the four vessel time series jointly using two latent temporal
processes: (i) a vessel-specific baseline trend
$\mathbf{f}^{(\mathrm{base})}(v,t)$ defined over the full study window, and
(ii) a vessel-specific disruption process
$\mathbf{f}^{(\mathrm{disrupt})}(v,t)$ present only for the starved vessels
$v\in\{1,2\}$. Let $v_n\in\{1,2,3,4\}$ denote the vessel associated with observation $n$. We fit
\begin{align*}
\mathbf{Y}_{\cdot n} &\sim  \text{Multinomial}(\lambda_n, \boldsymbol{\Pi}_{\cdot n}), \\ 
\boldsymbol{\Pi}_{\cdot n} &= \text{ALR}_{D}^{-1}(\mathbf{H}_{\cdot n}), \\ 
\mathbf{H}_{\cdot n} &\sim N(\mathbf{F}_{\cdot n}, \boldsymbol{\Sigma}), \\ 
  \mathbf{F}_{\cdot n} &= \sum_{v=1}^{4} I(v_n=v) \left[\mathbf{f}^{(\mathrm{base},v)}(t_n) + I(v\in\{1,2\}) \mathbf{f}^{(\mathrm{disrupt},v)}(t_n) \right],
 \\ 
\mathbf{f}^{(\text{base}, v)} &\sim \text{GP}(\boldsymbol{\Theta}^{(\text{base}, v)}, \boldsymbol{\Sigma}, \boldsymbol{\Gamma}^{(v)}  \odot\boldsymbol{\Gamma}^{(\text{base})}), v \in \{1,2,3,4\},\\
\mathbf{f}^{(\text{disrupt}, v)} &\sim \text{GP}(\boldsymbol{\Theta}^{(\text{disrupt}, v)}
, \boldsymbol{\Sigma}, \boldsymbol{\Gamma}^{(v)}  \odot \boldsymbol{\Gamma}^{(\text{disrupt})}), v \in \{1,2\},\\
\boldsymbol{\Sigma} &\sim \text{InvWishart}(\boldsymbol{\Xi}, \zeta).
\end{align*}
Here, $\boldsymbol{\Xi}$ is a fixed $(D-1)\times(D-1)$ matrix
with diagonal entries equal to 1 and off-diagonal entries equal to 0.4. Thus,
the same prior covariance is assigned to every pair of distinct ALR coordinates. We set $\zeta=D+5$. All prior mean
functions were set to zero. Continuous
covariates were standardized before fitting. No intercept was included, and
posterior draws were therefore not post-centered.

\paragraph{Vessel independence and kernel construction}

Although the four vessel trajectories are represented jointly within a single
latent process, the vessels are physically isolated and are therefore modeled
as independent by enforcing block-diagonal covariance across vessels. Let
$\boldsymbol{\Gamma}^{(v)}$ denote a vessel indicator kernel that equals one
for observations from the same vessel and zero for observations from different
vessels, and let $\odot$ denote the Hadamard product. The resulting baseline covariance has the form
\[
\boldsymbol{\Gamma}^{(v)}
\odot
\boldsymbol{\Gamma}^{(\mathrm{base})}
=
\begin{bmatrix}
\boldsymbol{\Gamma}^{(v=1,\mathrm{base})} & 0 & 0 & 0\\
0 & \boldsymbol{\Gamma}^{(v=2,\mathrm{base})} & 0 & 0\\
0 & 0 & \boldsymbol{\Gamma}^{(v=3,\mathrm{base})} & 0\\
0 & 0 & 0 & \boldsymbol{\Gamma}^{(v=4,\mathrm{base})}
\end{bmatrix},
\]
and the disruption covariance is block diagonal over the two starved vessels,
\[
\boldsymbol{\Gamma}^{(v)}
\odot
\boldsymbol{\Gamma}^{(\mathrm{disrupt})}
=
\begin{bmatrix}
\boldsymbol{\Gamma}^{(v=1,\mathrm{disrupt})} & 0\\
0 & \boldsymbol{\Gamma}^{(v=2,\mathrm{disrupt})}
\end{bmatrix}.
\]
Thus, each vessel has its own temporal trajectory, while covariance between
observations from different vessels is constrained to zero.

\paragraph{Temporal kernels}
We model baseline trends using a squared exponential kernel:
\[
\Gamma^{(\text{base})}(t_n,t_{n'}) =
\sigma_{\text{base}}^2 \exp\!\left\{- \frac{(t_n-t_{n'})^2}{2 \rho_{\text{base}}^2}\right\}.
\]
For the disruption effects, we employ a rational quadratic kernel set to zero prior to day 11. This specification reflects the assumption that the target variable exhibits varying degrees of smoothness and irregularities near or after the starvation period:

\[
\Gamma^{(\text{disrupt})}(t_n,t_n') = \sigma_{\text{disrupt}}^2 \left\{1+\frac{(t_n-t_n')^2}{2a \rho_{\text{disrupt}}^2}\right\}^{-a}I(t_n\geq 11\, \&\, t_n'\geq 11).
\]

\paragraph{Numerical regularization and back-sampling}
The gated component kernels above are positive semidefinite and have zero rows and columns outside their active vessel or time window. The analysis code adds $\epsilon\mathbf{I}_N$, with $\epsilon=10^{-10}$, to each component covariance after applying the vessel and time gates. The implemented covariance is therefore
\[
\widetilde{\boldsymbol{\Gamma}}^{(k)}
=\boldsymbol{\Gamma}^{(k)}+10^{-10}\mathbf{I}_N.
\]
The back-sampler uses these positive-definite matrices when inverting a component covariance and the sum of the remaining component covariances. Assigning the last component as a residual does not by itself remove the need for these inverses. At an inactive sample, the conditional prior covariance of that component is $10^{-10}\boldsymbol{\Sigma}$ rather than zero. Thus, the implemented disruption process approximates the zero-before-starvation constraint with a small positive variance; it is not exactly zero there. This kernel regularization is separate from any stabilization of the Laplace Hessian.

\begin{figure}[!t]
    \centering
    \includegraphics[width=\linewidth]{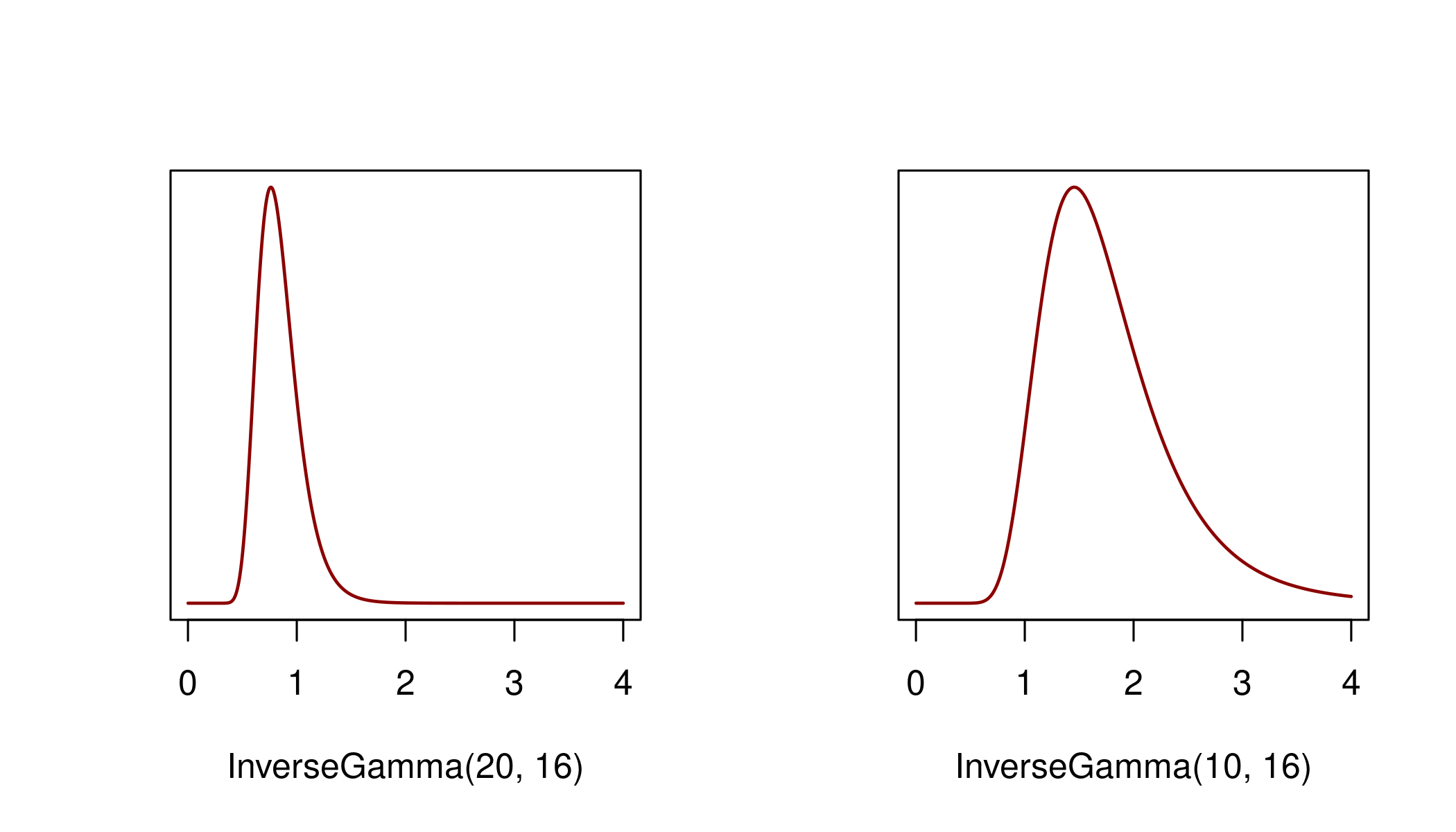}
    \caption{Base InverseGamma densities used to construct the kernel-hyperparameter regularization penalty for the artificial-gut analysis.}
    \label{s-fig:priordensity}
\end{figure}

We constructed the hyperparameter regularization penalty using the following inverse-gamma base densities for the length-scale parameters \((\rho_{\mathrm{base}},\rho_{\mathrm{disrupt}})\) and magnitude parameters \((\sigma_{\mathrm{base}},\sigma_{\mathrm{disrupt}})\):
\begin{align*}
\rho_{\text{base}} &\sim \text{InverseGamma}(10,16), &
\sigma_{\text{base}} &\sim \text{InverseGamma}(20,16), \\
\rho_{\text{disrupt}} &\sim \text{InverseGamma}(20,16), &
\sigma_{\text{disrupt}} &\sim \text{InverseGamma}(10,16).
\end{align*}
See Figure~\ref{s-fig:priordensity} for the corresponding density plots. Note that we fixed the $a$ parameter in the rational quadratic kernel, which determines the relative weighting of large-scale and small-scale variations, at a value of 2. This specification reflects our assumption that the model is constrained from learning distances that are significantly smaller or larger than the temporal distances among $t$. In other words, the regularization penalty discourages extremely small or large length scales. Let $\mathcal{K}=\{\rho_{\mathrm{base}},\rho_{\mathrm{disrupt}},\sigma_{\mathrm{base}},\sigma_{\mathrm{disrupt}}\}$. The penalized log marginal likelihood, with penalty weight $\kappa$, is given by
\begin{align*}
  \label{s-eq:laplace-approx-marginallikelihood_penalty1} 
  \ell_{\mathrm{pen}}(\Omega) &:= \log \int p(\mathbf{H},\mathbf{Y}\mid\Omega)\,d\mathbf{H} + \kappa\sum_{\theta\in\mathcal{K}}\log p(\theta)\\
  &\approx \frac{(D-1)N}{2}\log(2\pi) +  \log p(\widehat{\mathbf{H}}_{\Omega}, \mathbf{Y} \mid  \Omega)-\frac{1}{2} \log \left|\nabla^2\left\{\operatorname{vec}(\widehat{\mathbf{H}}_{\Omega})\right\}\right|\\
   &\quad + \kappa \times [\log p(\rho_{\text{base}})+ \log p(\rho_{\text{disrupt}}) +\log p(\sigma_{\text{base}})+ \log p(\sigma_{\text{disrupt}})].
\end{align*}

\begin{figure}[!t]
    \centering
    \includegraphics[width=\linewidth]{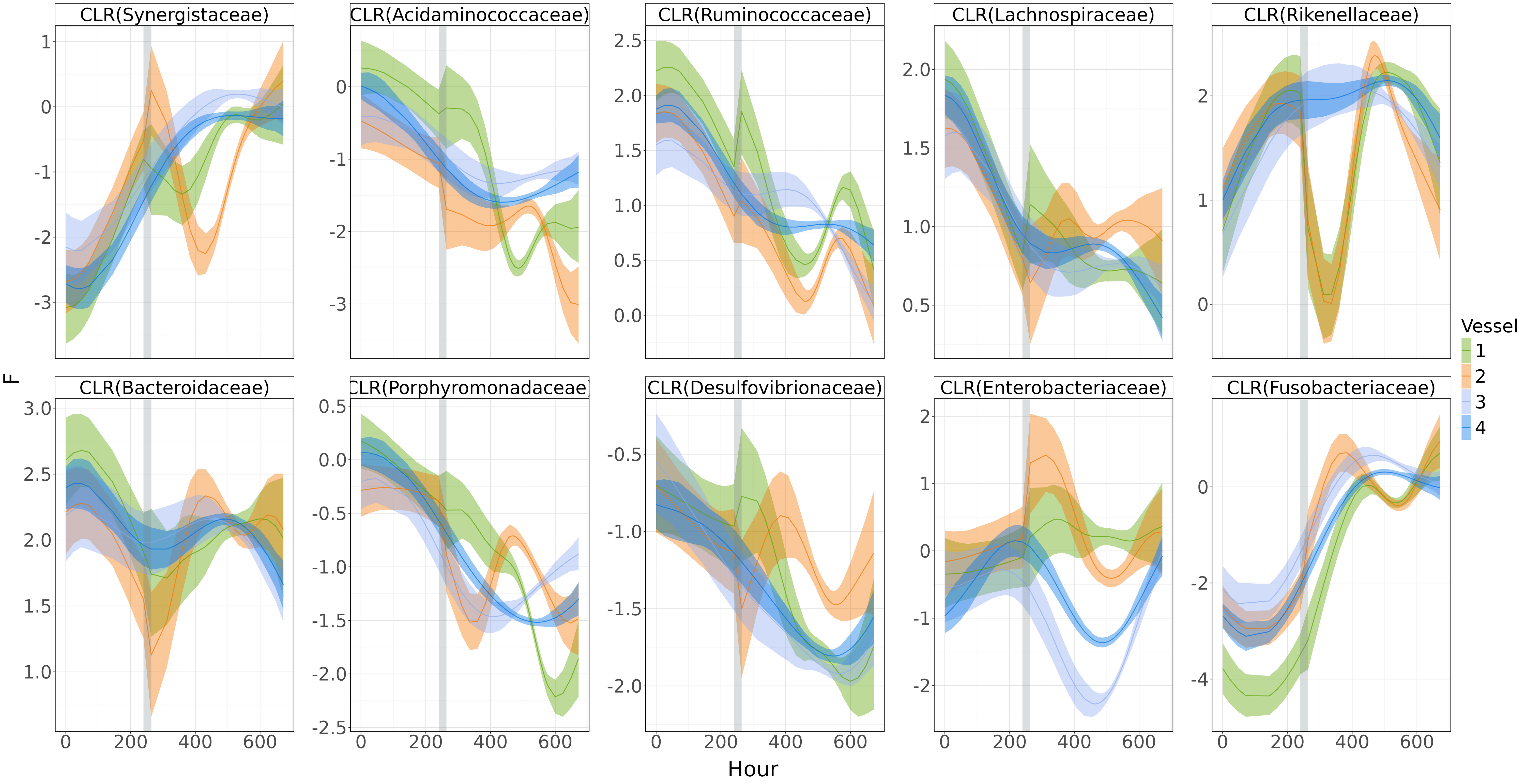}
    \caption{\textbf{MultiAddGPs Posterior for the Smoothed State \(\mathbf{F}\) of an Artificial Gut Study.} As described in the main text, \(\mathbf{F}\) was modeled as the additive combination of a slowly-varying temporal trend and a more variable yet short-lived starvation effect in Vessels 1 and 2 which started on Day 11. Posterior mean and 95\% credible intervals of \(\mathbf{F}\) are shown for each of the four artificial gut vessels. We plot these results with respect to the Centered Log-Ratio (CLR) Coordinates of each bacterial family. The starvation event, which occurred between Days 11 and 12 in Vessels 1 and 2, is highlighted in Gray.} 
    \label{s-fig:artificial-gut-cumulative}               
\end{figure}

\begin{figure}[!t]
     \centering
    \includegraphics[width=\linewidth]{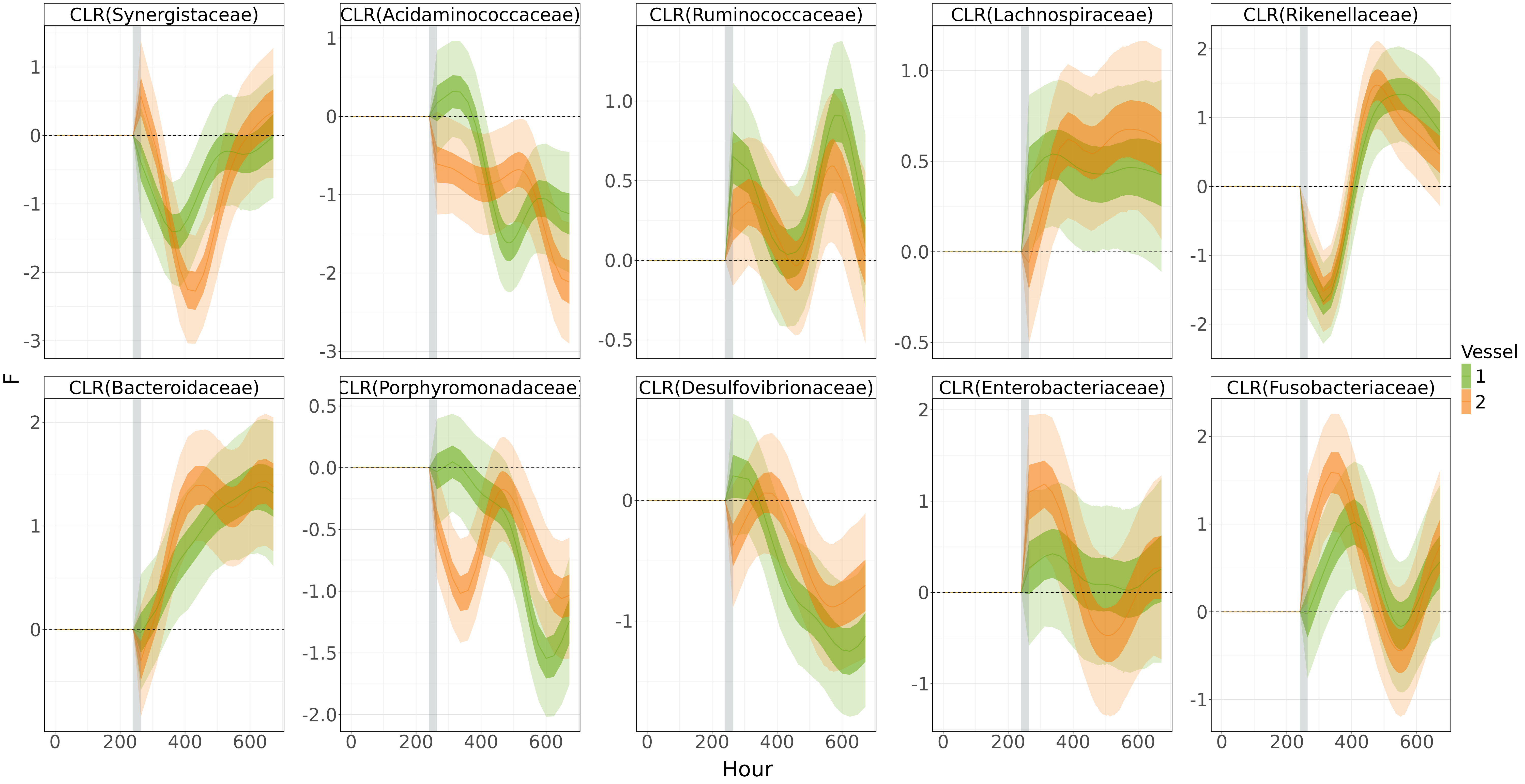}
    \caption{\textbf{MultiAddGPs Posterior for the Starvation effect ($\mathbf{F}^{\text{(disrupt)}}$) in Artificial Gut Study.} Posterior mean, 50\% and 95\% credible intervals are shown for all taxa.}
    \label{s-fig:artificial-effect}
\end{figure}

To aid in model identification, we also imposed the constraint $\sigma_{\text{base}} < \sigma_{\text{disrupt}}$, assuming greater variation is attributed to the starvation kernel following prior reports~\citep{s-silverman2018dynamic}. Finally, we set $\rho_{\text{base}} > \rho_{\text{disrupt}}$, reflecting the expectation that, in the absence of starvation, the base kernel should exhibit smoother and flatter trends. Note that we did not center the posterior samples at a mean of 0, as no intercept was included in the model.
We chose \(\kappa=120\) which was the smallest value that provided clear separation between the distinction between \(\mathbf{f}^{(\text{disrupt})}\) and \(\mathbf{f}^{(\text{base})}\). No perceptible change in estimated disruption effects was observed for values between 120 and 200. Posterior summaries of the cumulative latent trajectory $\mathbf{F}$ (long-term trend plus starvation effect) are shown in Figure~\ref{s-fig:artificial-gut-cumulative}, and posterior summaries of the isolated starvation effect are shown in Figure~\ref{s-fig:artificial-effect}.


\subsection{Human longitudinal microbiome data}
\label{s-supp-sec:long-data}

We applied MultiAddGPs to the daily longitudinal gut microbiome time series from \citet{s-david2014host}, focusing on Subject A. Sequence data were obtained from the European Nucleotide Archive (study accession ERP006059; Subject A nutritional metadata group accession SAMEG179160) and processed following \citet{s-david2014host}, except that we used DADA2~\citep{s-callahan2016dada2} for sequence-variant calling. We used DADA2 because it provides sequence-variant resolution and dataset-specific error modeling, improving reproducibility and taxonomic resolution. 
We then aggregated to the family level, yielding \(D=47\) bacterial families across \(N=329\) samples.

\paragraph{Model specification}
\label{s-supp-sec:long-data-model}

We modeled the data using two additive latent temporal components: a smooth general component, \(\mathbf{f}^{(\mathrm{general})}\), and a travel-associated component, \(\mathbf{f}^{(\mathrm{travel})}\). Specifically,
\begin{align*}
\mathbf{Y}_{\cdot n} &\sim \mathrm{Multinomial}(\lambda_n, \boldsymbol{\Pi}_{\cdot n}),\\
\boldsymbol{\Pi}_{\cdot n} &= \mathrm{ALR}_{D}^{-1}(\mathbf{H}_{\cdot n}),\\
\mathbf{H}_{\cdot n} &\sim N(\mathbf{F}_{\cdot n}, \boldsymbol{\Sigma}),\\
\mathbf{F}_{\cdot n} &= \mathbf{b}_0 + \mathbf{f}^{(\mathrm{general})}(t_n) + \mathbf{f}^{(\mathrm{travel})}(t_n),\\
\mathbf{f}^{(\mathrm{general})} &\sim \mathcal{GP}\!\left(\boldsymbol{\Theta}^{(\mathrm{general})}, \boldsymbol{\Sigma}, \boldsymbol{\Gamma}^{(\mathrm{general})}\right),\\
\mathbf{f}^{(\mathrm{travel})} &\sim \mathcal{GP}\!\left(\boldsymbol{\Theta}^{(\mathrm{travel})}, \boldsymbol{\Sigma}, \boldsymbol{\Gamma}^{(\mathrm{travel})}\right),\\
\boldsymbol{\Sigma} &\sim \mathrm{InvWishart}(\boldsymbol{\Xi}, \zeta).
\end{align*}
For the taxon covariance prior, we set $\zeta=D+100$ and \(\boldsymbol{\Xi}=(\zeta-D)\mathbf{G}\mathbf{G}^\top\), where \(\mathbf{G}=[\mathbf{I}_{D-1},-\mathbf{1}_{D-1}]\) is the additive log-ratio contrast matrix. For the general component, time was standardized to mean zero and variance one before fitting. All kernel mean functions were set to zero. Posterior samples were centered to have mean zero.

\paragraph{Temporal kernels and warped travel component}
\label{s-supp-sec:long-data-kernel}

Both latent components were modeled with squared exponential kernels, with the travel component defined on a warped time axis:
\begin{align*}
\Gamma^{(\mathrm{general})}(t_n,t_{n'})
&=
\sigma_{\mathrm{general}}^2
\exp\!\left\{-\frac{(t_n-t_{n'})^2}{2\rho_{\mathrm{general}}^2}\right\},\\
\Gamma^{(\mathrm{travel})}(t_n,t_{n'})
&=
\sigma_{\mathrm{travel}}^2
\exp\!\left[-\frac{\{w(t_n)-w(t_{n'})\}^2}{2\rho_{\mathrm{travel}}^2}\right].
\end{align*}

Here, $w_n=w(t_n)$ is a transformed skewed Gaussian of input $t_n$, designed to capture the non-stationary nature of the travel intervention. The travel-warping function was constructed on the original day scale. Specifically:
\begin{align*}
z_n &= \frac{t_n - m}{s},\\
w_n &= \exp\left(-\frac{1}{2}z_n^2\right) \times \frac{1}{1+ \exp(-\alpha z_n)},
\end{align*}
where $m$ determines the peak location of the distribution, $s$ governs its width, and $\alpha$ is the skew parameter.
As in the simulation study (Section~\ref{s-subsec:sim-lgpr-gam}), the resulting $w_n$ values were normalized to $[-1,1]$ across the analyzed collection days before use in the travel kernel.
When $\alpha=0$, the function reduces to a standard Gaussian; a positive
value of $\alpha$ induces a heavier right tail, while a negative value
produces a heavier left tail. Here we set $m=92$, $s=20$, and $\alpha=2$. These choices effectively restrict the travel-related changes to a two-month window after the travel date, ensuring that the model captures the local, non-stationary fluctuations associated with travel without introducing undue influence at other time points.

\begin{figure}[!t]
    \centering
    \includegraphics[width=\linewidth]{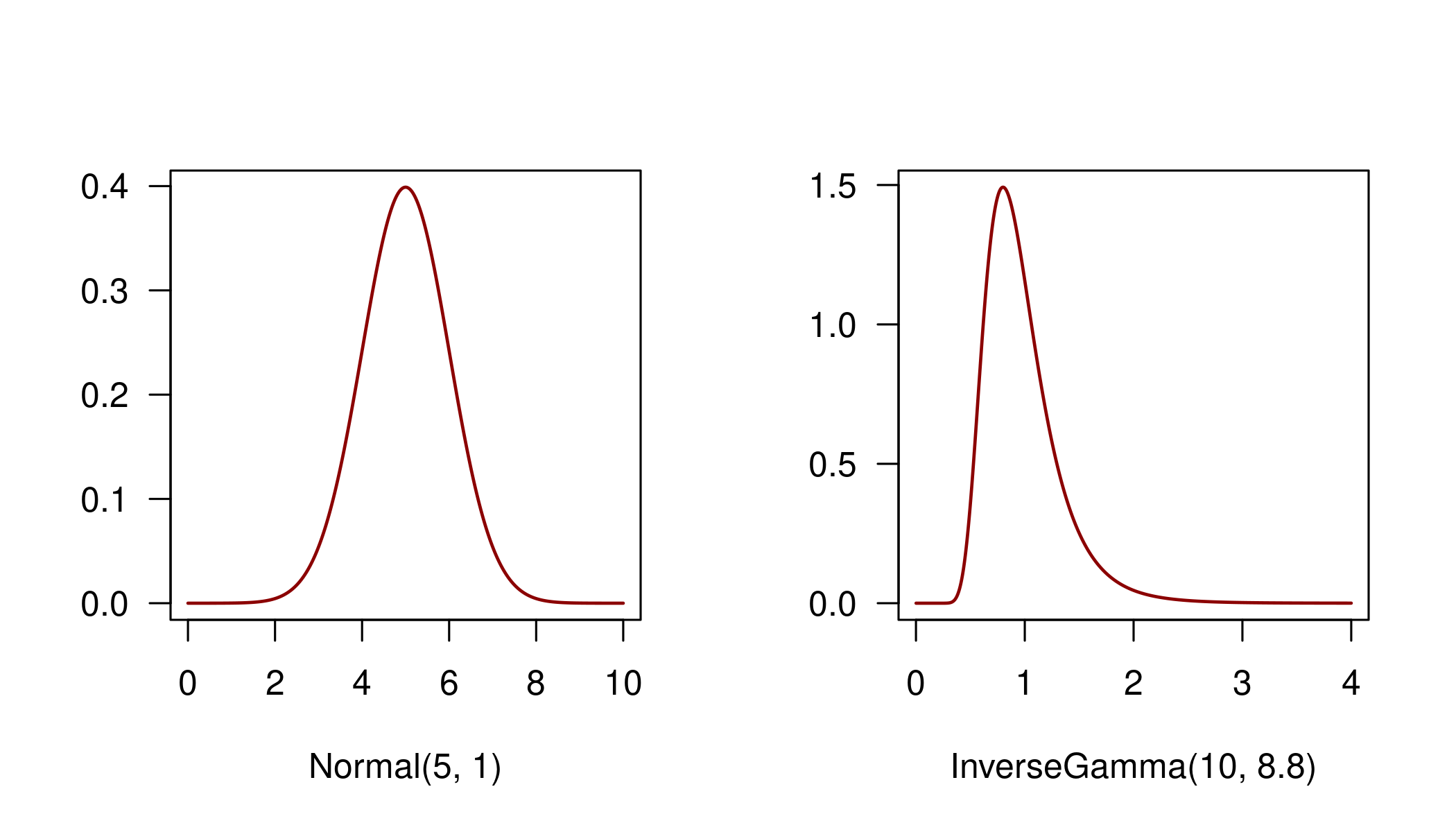}
    \caption{Base Normal and InverseGamma densities used to construct the kernel-hyperparameter regularization penalty for the human longitudinal microbiome analysis.}
    \label{s-fig:prior_longitudinal}
\end{figure}

We constructed the hyperparameter regularization penalty using the following base densities:
\begin{align*} \sigma_{\text{general}}, \sigma_{\text{travel}}, \rho_{\text{travel}} &\sim \mathcal{N}(5,1),\\
\rho_{\text{general}} &\sim \text{InverseGamma}(10,8.8). \end{align*}
Here, $\mathrm{InverseGamma}(a,b)$ uses shape $a$ and scale $b$, with density $b^a x^{-a-1}\exp(-b/x)/\Gamma(a)$ for $x>0$. The length-scale penalty therefore has mode $8.8/(10+1)=0.8$ on the standardized time scale. The corresponding base densities are shown in Figure~\ref{s-fig:prior_longitudinal}. The separation of background and travel-associated components was motivated by the months of overall stability and the marked, reversible travel-associated shift reported by \citet{s-david2014host}. These densities define the regularization penalty for the kernel hyperparameters during optimization. Let $\mathcal{K}=\{\rho_{\mathrm{general}},\rho_{\mathrm{travel}},\sigma_{\mathrm{general}},\sigma_{\mathrm{travel}}\}$. The penalized log marginal likelihood is given by
\begin{align*}
  \ell_{\mathrm{pen}}(\Omega) &:= \log \int p(\mathbf{H},\mathbf{Y}\mid\Omega)\,d\mathbf{H} + \kappa\sum_{\theta\in\mathcal{K}}\log p(\theta)\\
  &\approx \frac{(D-1)N}{2}\log(2\pi) +  \log p(\widehat{\mathbf{H}}_{\Omega}, \mathbf{Y} \mid  \Omega)-\frac{1}{2} \log \left|\nabla^2\left\{\operatorname{vec}(\widehat{\mathbf{H}}_{\Omega})\right\}\right|\\
   &\quad + \kappa \times [\log p(\rho_{\text{general}})+ \log p(\rho_{\text{travel}}) +\log p(\sigma_{\text{general}})+ \log p(\sigma_{\text{travel}})].
\end{align*}

The densities $p(\theta)$ define the regularization penalty. To help distinguish the two temporal processes, we imposed the constraints $\sigma_{\text{travel}} < \sigma_{\text{general}}$ and $\rho_{\text{general}} < \rho_{\text{travel}}$, reflecting the expectation of a localized, smooth travel response on time with a smaller kernel amplitude than the general component. Here, $\rho_{\mathrm{general}}$ is defined on standardized time and $\rho_{\mathrm{travel}}$ on the warped input normalized to $[-1,1]$, so their numerical ordering does not directly imply an ordering of smoothness in calendar time.
As in the artificial-gut analysis (Section~\ref{s-subsec:artificial_data}),
we used a penalized marginal likelihood and set $\kappa=100$.

\paragraph{Data-adaptive balance analysis of the isolated travel effect}
\label{s-supp-sec:long-data-principal-balance}

We summarized posterior samples of the isolated travel component $\mathbf{F}^{(\mathrm{travel})}$ using principal balances. We represented the samples used for this analysis as the collection $\{\boldsymbol{\Lambda}^{(\mathrm{travel},s)}\}_{s=1}^{S}$, where $\boldsymbol{\Lambda}^{(\mathrm{travel},s)}\in\mathbb{R}^{D\times N}$ is the matrix of values for $D$ taxa across $N$ time points in posterior draw $s$. For each draw, we computed the taxon covariance across time, $\mathbf{W}^{(s)}=\mathrm{Cov}(\boldsymbol{\Lambda}^{(\mathrm{travel},s)})$, and converted it to a variation matrix $\mathbf{T}^{(s)}$ with entries $T^{(s)}_{ij}=W^{(s)}_{ii}+W^{(s)}_{jj}-2W^{(s)}_{ij}$. We applied Ward clustering to $\mathbf{T}^{(s)}$ to obtain a tree for each draw, then formed a majority-rule ($p=0.5$) consensus tree. We randomly resolved multifurcations in the consensus tree to obtain a binary tree, then converted its sequential binary partition into an orthonormal ILR contrast matrix $\mathbf{V}\in\mathbb{R}^{D\times(D-1)}$ (balances $w_1,\dots,w_{D-1}$). Column $k$ of $\mathbf{V}$ contains the ILR weights for balance $w_k$ across the $D$ taxa, with positive entries corresponding to the numerator group and negative entries to the denominator group.

\begin{figure}
    \centering
    \includegraphics[width=\linewidth,height=0.85\textheight,keepaspectratio]{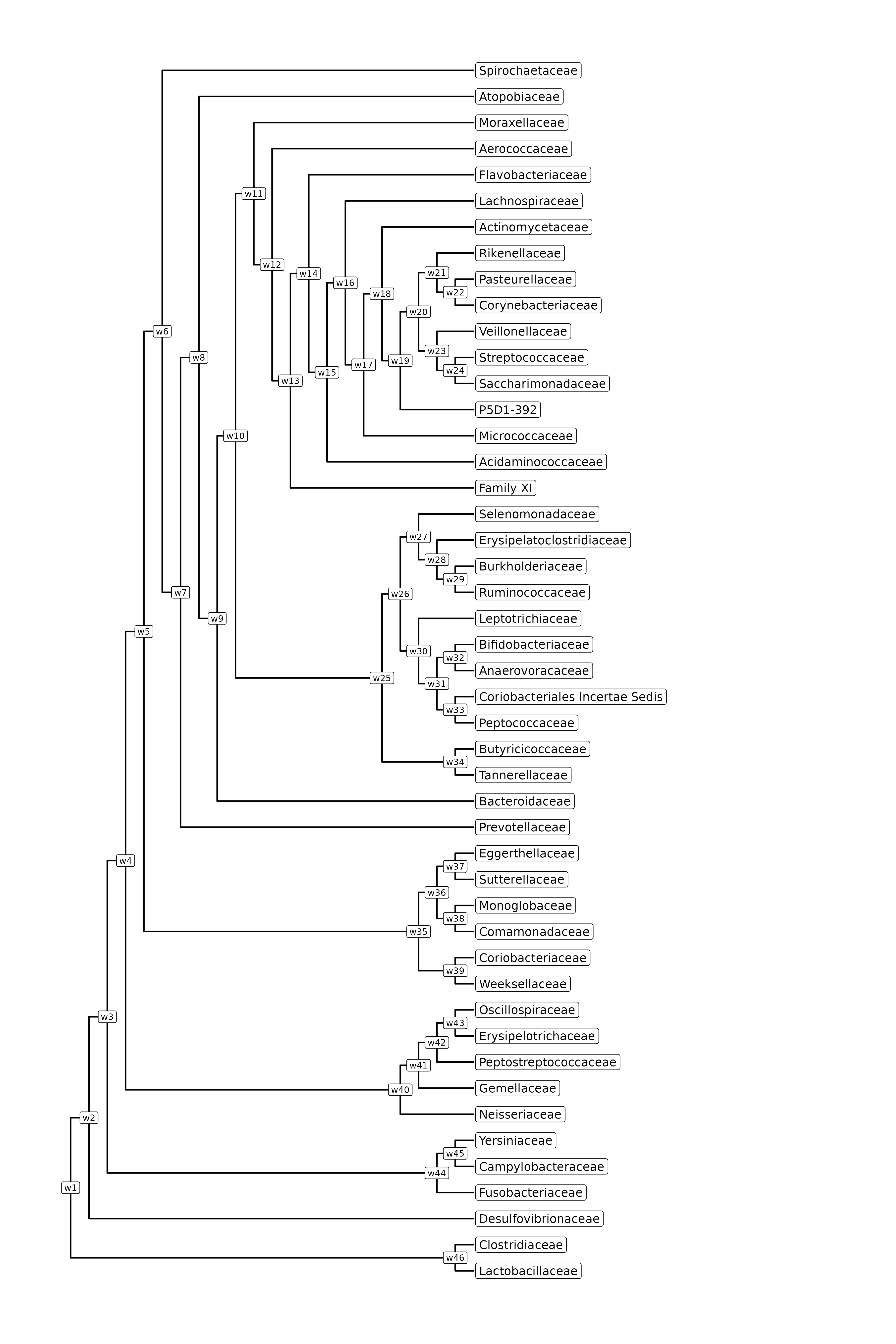}
    \caption{The consensus tree created by Ward clustering of bacterial families with Ward balances 
    (all balances) is depicted. }
    \label{s-fig:balance_tree}
\end{figure}

We projected each posterior draw into balance coordinates using $\mathbf{U}^{(s)}=\mathbf{V}^{\top}\boldsymbol{\Lambda}^{(\mathrm{travel},s)}\in\mathbb{R}^{(D-1)\times N}$. The row vector $\mathbf{U}_{k\cdot}^{(s)}$ gives the time series for balance $w_k$ in draw $s$. To identify dominant balances, we computed the variance over time for each balance and draw, $v_k^{(s)}=\mathrm{Var}(\mathbf{U}_{k\cdot}^{(s)})$, and ranked balances by the posterior mean $\overline{v}_k=(1/S)\sum_{s=1}^{S}v_k^{(s)}$. We report the highest-ranked balance in Figure~\ref{m-fig:lifestyle_data_principalbalance} of the main text; its taxon weights are given by the corresponding column of $\mathbf{V}$. A full balance tree is provided in Figure~\ref{s-fig:balance_tree}.

To assess sensitivity to the binary resolution of the consensus tree, we examined 50 resolutions in total, including the reference resolution, while holding the posterior draws and draw-specific Ward trees fixed. In all 50, the highest-ranked balance placed \textit{Fusobacteriaceae}, \textit{Yersiniaceae}, and \textit{Campylobacteraceae} on the same side; in 40, these three families formed one side of one of the two highest-ranked balances by themselves. These results support the three-family grouping across resolutions, although balance labels and opposing taxon groups varied.

\begin{figure}[!t]
    \centering
    \includegraphics[width=\linewidth]{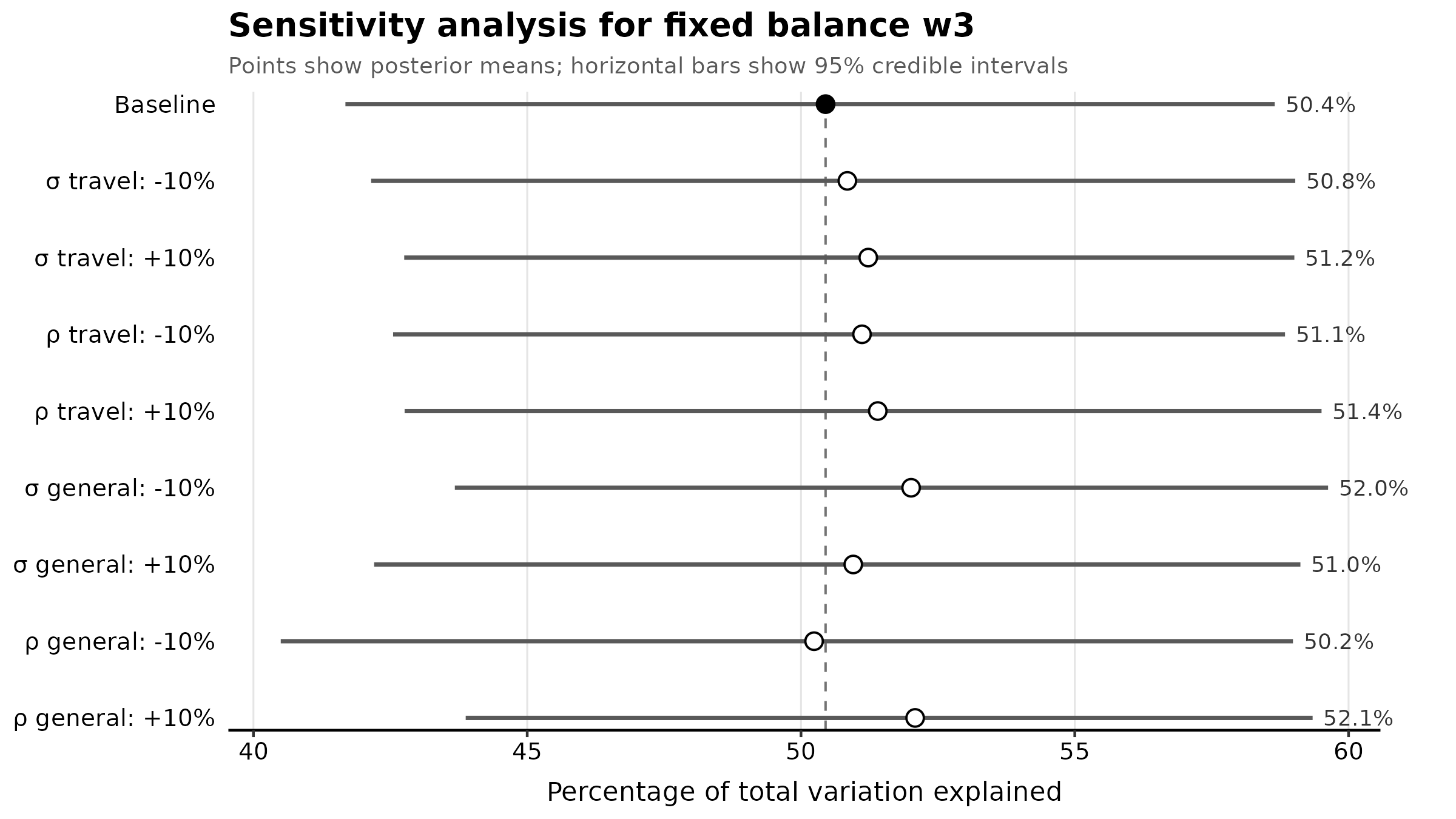}
    \caption{Posterior estimates for the percentage of total travel-associated variation
    explained by balance $w_3$ were not sensitive to modification of the hyperparameters.}
    \label{s-fig:sen_w3}
\end{figure}

Under the reference resolution, $w_3$ explained 50.4\% (95\% credible interval 41.7\%--58.7\%) of posterior travel-associated temporal variation. To assess hyperparameter sensitivity, we refit the model under eight one-at-a-time $\pm10\%$ changes to $\sigma_{\mathrm{travel}}$, $\rho_{\mathrm{travel}}$, $\sigma_{\mathrm{general}}$, and $\rho_{\mathrm{general}}$, holding the other hyperparameters fixed. We applied the reference direction of $w_3$ to each refit and recomputed its share of travel-associated variation. Across these refits, the mean share ranged from 50.2\% to 52.1\%. Posterior means and 95\% credible intervals appear in Figure~\ref{s-fig:sen_w3}.

\end{document}